\documentclass[10pt]{article}

\usepackage{fullpage}
\usepackage{setspace}
\usepackage{parskip}
\usepackage{titlesec}
\usepackage{xcolor}
\usepackage{lineno}
\usepackage[export]{adjustbox}

\usepackage[version=4]{mhchem}
\usepackage{siunitx}
\DeclareSIUnit\Molar{M}

\usepackage{mathtools}% http://ctan.org/pkg/mathtools

\PassOptionsToPackage{hyphens}{url}
\usepackage[colorlinks = true,
            linkcolor = blue,
            urlcolor  = blue,
            citecolor = blue,
            anchorcolor = blue]{hyperref}
\usepackage{etoolbox}

\renewenvironment{abstract}
  {{\bfseries\noindent{\abstractname}\par\nobreak}\normalsize}
  {\bigskip}

\titlespacing{\section}{0pt}{*3}{*1}
\titlespacing{\subsection}{0pt}{*2}{*0.5}
\titlespacing{\subsubsection}{0pt}{*1.5}{0pt}

\usepackage{authblk}

\usepackage{graphicx}
\usepackage[space]{grffile}
\usepackage{latexsym}
\usepackage{textcomp}
\usepackage{longtable}
\usepackage{tabulary}
\usepackage{booktabs,array,multirow}
\usepackage{amsfonts,amsmath,amssymb}
\providecommand\citet{\cite}
\providecommand\cite{\cite}

\newif\iflatexml\latexmlfalse

\AtBeginDocument{\DeclareGraphicsExtensions{.pdf,.PDF,.eps,.EPS,.png,.PNG,.tif,.TIF,.jpg,.JPG,.jpeg,.JPEG}}

\usepackage[utf8]{inputenc}
\usepackage[english]{babel}

\begin{document}

\doublespacing

\title{BCI learning induces core-periphery reorganization in M/EEG multiplex brain networks}

\author[a,b,*]{Marie-Constance Corsi}
\author[c]{Mario Chavez} 
\author[d]{Denis Schwartz}
\author[d]{Nathalie George}
\author[d]{Laurent Hugueville}
\author[e]{Ari E. Kahn}
\author[b]{Sophie Dupont}
\author[e,f,g,h,i,j]{Danielle S. Bassett}
\author[a,b,*]{Fabrizio De Vico Fallani}

\affil[a]{Inria Paris, Aramis project-team, F-75013, Paris, France}
\affil[b]{Institut du Cerveau et de la Moelle Epini\`ere, ICM, Inserm, U 1127, CNRS, UMR 7225, Sorbonne Université, F-75013, Paris, France}
\affil[c]{CNRS, UMR 7225, F-75013, Paris, France}
\affil[d]{Institut du Cerveau et de la Moelle Epini\`ere, ICM, Inserm U 1127, CNRS UMR 7225, Sorbonne Université, Ecole Normale Supérieure, ENS, Centre MEG-EEG, F-75013, Paris, France}
\affil[e]{Department of Bioengineering, School of Engineering and Applied Science, University of Pennsylvania, Philadelphia, PA 19104, USA}
\affil[f]{Department of Neurology, Perelman School of Medicine, University of Pennsylvania, Philadelphia, PA 19104, USA}
\affil[g]{Department of Physics and Astronomy, College of Arts and Sciences, University of Pennsylvania, Philadelphia, PA 19104, USA}
\affil[h]{Department of Electrical and Systems Engineering, School of Engineering and Applied Science, University of Pennsylvania, Philadelphia, PA 19104, USA}
\affil[i]{Department of Psychiatry, Perelman School of Medicine, University of Pennsylvania, Philadelphia, PA 19104 USA}
\affil[j]{Santa Fe Institute, Santa Fe, NM 87501 USA}

\affil[*]{Corresponding authors: 
Marie-Constance Corsi marie.constance.corsi@gmail.com;
Fabrizio De Vico Fallani fabrizio.devicofallani@gmail.com}

\vspace{-1em}

%\date{\today}

\begingroup
\let\center\flushleft
\let\endcenter\endflushleft
\maketitle
\endgroup

\selectlanguage{english}

% \textbf{Highlights}:
% \begin{enumerate}
%     \item Brain networks elicit opposite trends between \alpha and \beta frequency ranges
%     \begin{enumerate}
%         \item \alpha:  increase of relative coreness (from negative to positive values), involving somatosensory and motor planning areas
%         \item \beta: decrease of relative coreness, in regions involved in visual process, working memory and motor planning
%     \end{enumerate}
%     \item At the end of the BCI training: equal contribution of E/MEG with time to maximize subjects’ mental state discrimination \&  captures globally the same task-related processes
%     \item Brain network properties resulting from the E/MEG integration correlated with future BCI score in \alpha2 : positively in  areas involved in decision making \& somatosensory, negatively in visual processing areas
% \end{enumerate}

\begin{abstract}
\textbf{Objective:} Brain-computer interfaces (BCIs) constitute a promising tool for communication and control. However, mastering non-invasive closed-loop systems remains a learned skill that is difficult to develop for a non-negligible proportion of users. The involved learning process induces neural changes associated with a brain network reorganization that remains poorly understood. \\
\textbf{Approach:} To address this inter-subject variability, we adopted a multilayer approach to integrate brain network properties from electroencephalographic (EEG) and magnetoencephalographic (MEG) data resulting from a four-session BCI training program followed by a group of healthy subjects. Our method gives access to the contribution of each layer to multilayer network that tends to be equal with time. \\
\textbf{Main results:} We show that regardless the chosen modality, a progressive increase in the integration of somatosensory areas in the $\alpha$ band was paralleled by a decrease of the integration of visual processing and working memory areas in the $\beta$ band. Notably, only brain network properties in multilayer network correlated with future BCI scores in the $\alpha_2$ band: positively in somatosensory and decision-making related areas and negatively in associative areas. \\
\textbf{Significance:} Our findings cast new light on neural processes underlying BCI training. Integrating multimodal brain network properties provides new information that correlates with behavioral performance and could be considered as a potential marker of BCI learning.
\end{abstract}

%\newpage

\section*{Introduction}
Learning is a complex phenomenon that can be characterized by changes in regional associations and therefore in brain network organization \cite{reddy_brain_2018}. Changes following learning have been revealed in language \cite{deng_resting-state_2016, sheppard_large-scale_2012} and in motor skill acquisition with resting-state fMRI recordings \cite{sami_graph_2013, taubert_long-term_2011}. In the case of motor learning, studies that focus on functional connectivity have demonstrated changes induced by skill acquisition 
\cite{ge_motor_2015, mcdougle_taking_2016,rizzolatti_cortical_2001, katiuscia_reorganization_2009, taubert_long-term_2011}.
From a network perspective, a large number of metrics characterizing network properties have been considered to capture the process of motor acquisition. In Ref. \cite{heitger_motor_2012}, the motor performance improvement was associated with an increase of clustering coefficients, a higher number of network connections, an increased connection strength and shorter communication distances. Another approach consists of using a single metric that measures subnetwork segregation: modularity \cite{sporns_modular_2016}, already used as a marker of brain plasticity \cite{gallen_brain_2019} and motor learning \cite{bassett_dynamic_2011}. Motor skill acquisition induced an autonomy of sensorimotor systems and individual differences in the amount of learning could be predicted by the release of cognitive control hubs in frontal and cingulate cortices \cite{bassett_learning-induced_2015}.

Mastering non-invasive closed-loop systems is a learned skill that requires several training sessions to achieve control of the device. Recent studies suggest that the involved learning process is analogous to cognitive or motor skill acquisition in the case of BCI \cite{hiremath_brain_2015}. It may induce behavioral modifications and neural changes within trained brain circuits in neurofeedback that last for several months after training \cite{sitaram_closed-loop_2016}. Changes at the neuronal level, during the learning process, have also been observed and simulated \cite{ito_self-reorganization_2020}. The recruitment of areas beyond those targeted by BCI has been observed during skill acquisition \cite{orsborn_parsing_2017, wander_distributed_2013}, and a decrease in the global efficiency index in the higher-beta frequency range with the practice of MI \cite{pichiorri_sensorimotor_2011} suggests the involvement of a distributed core of brain areas while learning. 
From a theoretical perspective, the existence of a core, a group of tightly connected nodes, surrounded by a poorly connected periphery is crucial for the integration of information between remote network components \cite{borgatti_models_2000,girvan_community_2002}. Previous studies have demonstrated the utility of using multilayer models of networks \cite{de_domenico_mathematical_2013, boccaletti_structure_2014} to study the relationship between structure and function in the human brain. The identification of core-periphery structures in brain networks can be significantly enriched by adding multiple levels of connectivity \cite{battiston_multilayer_2017, battiston_federico_multiplex_2018}. In particular, combining multifrequency or multimodal neuroimaging data from a network perspective can reveal higher-order topological properties that cannot be detected by simple single-layer network approaches \cite{de_domenico_mapping_2016, de_domenico_multilayer_2017, tewarie_integrating_2016, battiston_multilayer_2017, guillon_loss_2017, buldu_frequency-based_2018, guillon_disrupted_2019}.
Magnetoencephalography (MEG) and electroencephalography (EEG) are complementary in terms of 
sensitivity towards source depths and conductivity, but also in terms of dipole orientation detection  \cite{cuffin_comparison_1979, geisler_surface_1961, delucchi_scalp_1962, hamalainen_meg_theory_1993,wood_electrical_1985,sharon_advantage_2007}. 
As a result, their combination could provide valuable information, and has proven to enhance subjects' mental state discrimination in BCI \cite{corsi_integrating_2018}.

On the above mentioned elements, we hypothesized that integrating information from EEG and MEG data, allow a better description of the core-periphery changes occurring during a motor imagery-based BCI training in a group of healthy subjects. Such an enriched description could reveal fresh insights into learning processes that are difficult to observe at the single layer level and eventually improve the prediction of future BCI performance.

\section*{Materials and Methods}
\subsection*{Participants and experiment}
We included twenty healthy, and BCI naive, subjects (aged 27.5 $\pm$ 4.0~years, 12~men). All right-handed, they participated in a 4 session-based BCI training during two weeks. According to the declaration of Helsinki, a written informed consent was obtained from subjects after explanation of the study, which was approved by the ethical committee CPP-IDF-VI of Paris. 
The EEG-based BCI consisted of a two-target box task~\cite{wolpaw_wadsworth_2003}. The subjects were instructed to control the vertical position of a moving cursor by modulating the neural activity in the \(\alpha\) {[}8-12 Hz{]} and/or \(\beta\) {[}14-29 Hz{]} frequency bands.
Each session was divided into two phases: 
\begin{enumerate}
	\item The training phase consisted of five consecutive runs, of 32 trials each, without any feedback. For a given trial, the first second consisted of the inter-stimulus interval (ISI) followed by five seconds of target presentation. To elicit the (EEG electrodes; frequency bins) couples that best discriminate the subjects' mental state over the left motor area and within the mu-beta frequency ranges, we computed contrast maps that relied on the R-square metric \cite{schalk_bci2000:_2004}. 	
	\item The testing phase consisted of six runs, of 32 trials each, with a cursor feedback. Similarly to the training phase, for a given trial, we had one second of ISI, while the target was presented throughout the subsequent five seconds. The visual feedback, displayed from $t = 3 s$ to $t = 6 s$, consists of a moving cursor. The features, i.e. power spectra estimated at the (EEG electrodes; frequency bins) couples selected during the training, were classified by using the Linear Discriminant Analysis method. All the results presented in the following sections relied on the analysis performed on the testing data.
\end{enumerate} 
To perform the experiments, we used a 74~EEG-channel system, with Ag/AgCl passive sensors (Easycap, Germany) placed according to the standard 10-10 montage. The reference was located at the mastoids and the ground electrode was placed at the left scapula. We kept the impedances lower than 20~kOhms. The MEG system consisted of 102 magnetometers and 204 gradiometers (Elekta Neuromag TRIUX MEG system). E/MEG registrations were performed simultaneously in a magnetic shielded room with a sampling frequency of 1~kHz and a bandwidth of 0.01-300~Hz. The subjects were seated with palms facing upward in front of a 90 cm-distant screen.
To ensure that no forearm movements were performed, experts visually inspected electromyogram (EMG) signals recorded from the subject's right arm during the experiment. 
During the sessions, BCI feedback relied on EEG signals transmitted to the BCI2000 toolbox~\cite{schalk_bci2000:_2004}~via the Fieldtrip buffer~\cite{oostenveld_fieldtrip:_2010}.
Individual T1 sequences have been obtained by using a 3T Siemens Magnetom PRISMA after the fourth session to ensure accurate head models \cite{gross_good_2013}. These registrations consisted of a 15 minute-resting-state task. A preprocessing of the images was performed via the FreeSurfer toolbox \cite{fischl_freesurfer_2012} and directly imported (15002 vertices) to the Brainstorm toolbox. 
To provide co-registration with the anatomical MRI, we digitized the location of the EEG electrodes and three landmarks (nasion, left and righ pre-auricular points) with the FastTrak 3D digitizer (Polhemus, Inc., VT, USA). These locations were aligned with the MRI using the Brainstorm toolbox \cite{tadel_brainstorm:_2011}. A more detailed description of the experiments is proposed in Ref. \cite{corsi_functional_2020}.

\subsection*{Data analysis}
\subsubsection*{M/EEG processing}
After a first preprocessing step that consisted of an application of the temporal extension of the Signal Space Separation (tSSS) to MEG signals to remove environmental noise \cite{taulu_spatiotemporal_2006}, M/EEG data were downsampled to 250~Hz and processed via the Independent Component Analysis \cite{bell_information-maximization_1995,oostenveld_fieldtrip:_2010} to remove ocular and cardiac artifacts. Then, data were segmented into 7s-epochs, corresponding to the trial length. \\
Source reconstruction was performed by applying the Boundary Element Method \cite{fuchs_boundary_2001, gramfort_openmeeg:_2010} to obtain the individual head model, followed by the estimation of the sources with the weighted Minimum Norm Estimate \cite{fuchs_linear_1999,lin_assessing_2006,gramfort_mne_2014, tadel_brainstorm:_2011}. A more detailed description of the applied preprocessing steps is proposed in Ref. \cite{corsi_functional_2020}. \\
To compute the power spectra within the individual anatomical space, we used the Welch method. A time window of 1~s and a window overlap ratio of 50~\% was applied during the feedback period (i.e. from t~=~3~s to t~=~6~s) to obtain the cross-spectral estimation for each trial, session, and subject. Then, for each region of interest (ROI) from the Destrieux atlas \cite{destrieux_automatic_2010}, we took into account the first principal component of the power spectra computed over the dipoles. For each layer (or modality here) $l$ and frequency band $f$, we estimated the functional connectivity networks by computing the imaginary coherence between each pair of ROIs ($N=148$) \cite{sekihara_removal_2011}, resulting in 148 x 148 adjacency matrices $A_{l,f}$. 

%\clearpage

\subsubsection*{Network analysis and statistics}
Similarly to Refs \cite{battiston_federico_multiplex_2018, guillon_disrupted_2019}, to obtain the multilayer or multiplex brain networks $M_{f}$ for a given frequency band $f$ from the adjacency matrices $A_{l,f}$, we aligned the EEG and MEG connectivity networks as follows:
\begin{equation}
  M_{f}={A_{l,f}, \forall l \in \{EEG, MEG\}}, \label{eq:1}
\end{equation}
To study properties associated with a core-periphery organization, for a given layer (i.e. modality here), we filtered the associated adjacency matrix $A_{l,f}$ to keep the strongest weights by applying a broad range of thresholds corresponding to the average node degree $ k =1 $ to $k = N-1$. 
For each threshold $ k $, to determine whether a node $i $ belongs to the core, we computed the multiplex core-periphery of the filtered network by calculating its richness defined as follows:
\begin{equation}
  \mu_{i}=\sum_{l=1}^{L} c^{l}s_{i}^{l}, \label{eq:2}
\end{equation}
where $L$ corresponds to the number of layers ($L=2$), $ s_{i}^{l}$ corresponds to the strength of the node $i$ in the l-th layer (i.e. the sum of the i-th row of the matrix $A_{l,f}$), and $ c^{l}$ corresponds to the l-th component of the vector $c$ that represents the contribution of each layer (ranging from 0 to 1).
To take into account only the links of node i that are associated with nodes of higher richness, we decomposed the richness function as follows: $ s^{l}=s^{l-}+s^{l+}$. The richness of nodes linked to richer nodes can be defined as:
\begin{equation}
  \mu_{i+}=\sum_{l=1}^{L} c^{l}s_{i}^{l+}. \label{eq:3}
\end{equation}
We finally computed the multiplex coreness \cite{battiston_federico_multiplex_2018} $C_{i}$ of each node $i $, independently from any other consideration, by determining the number of times the node $i$ belongs to the core over all the $ k $ tested thresholds, as follows:
\begin{equation}
  C_{i}=\frac{1}{N-1} \sum_{k=1}^{N-1} \delta_{i}^{k}, \label{eq:4}
\end{equation}
where $ \delta_{i}^{k} =1$ if node $i$ belongs to the core for the threshold $k$, and $0 $ otherwise.
To obtain the coreness associated with a specific layer, one can simply modify the vector $ c $ in equation \ref{eq:3} so that the component not related to the given modality is equal to zero.
For each subject, session and frequency band, we optimized the choice of the components of the vector $c$ by using the Particles Swarm Optimization and Statistical Analysis (PSO) algorithm \cite{guillon_disrupted_2019, kennedy_particle_1995}. In our case, the Fisher's criterion $F(c)$, chosen to maximized the difference between the conditions, was defined as follows:
\begin{equation}
  F(c)=\frac{(<C^{MI}(c)>-<C^{rest}(c)>)^2}{(s^{MI})^2+(s^{rest})^2}, \label{eq:5}
\end{equation}
where $<C^{cond}(c)>$ is the averaged coreness computed over the nodes $i$ in the condition  $cond$ and
\begin{equation}
  (s^{cond})^2=\sum_{i \in \{1..N\}} (C_{i}^{cond}(c)-<C^{cond}(c)>)^2, \label{eq:6}
\end{equation}

where $C_{i}^{cond}$ corresponds to the coreness computed in node $i$ in the condition $cond$.

To study the variation of coreness between conditions, we defined the relative coreness ($\Delta$C) as $\Delta$C=$C^{MI}-C^{Rest}$.
To compute the multiplex core-periphery properties, we used the Brain Connectivity Toolbox \cite{rubinov_complex_2010} and the Matlab code available at \url{https://github.com/brain-network/bnt}.

To take into account the subjects' specificity, we used customized definitions of the~\(\alpha\) and~\(\beta\) bands~\cite{klimesch_eeg_1999}, that rely on the Individual Alpha Frequency (IAF)~\cite{pichiorri_brain-computer_2015}, obtained from a 3-minute resting state recording. The \(\alpha_1\) ranges from IAF~-~2~Hz to IAF,~\(\alpha_2\) from IAF to IAF~+~2~Hz,~\(\beta_1\) from IAF~+~2~Hz to IAF~+~11~Hz and~\(\beta_2\) from IAF~+~11~Hz to IAF~+~20~Hz.  Preliminary results did not show particularly significant efects in \(\theta\) and low \(\gamma\)
bands. Therefore, only results obtained within the \(\alpha\) and \(\beta\) frequency bands are presented here. \\
After plotting quantile-quantile plots and performing the Shapiro-Wilk test \cite{shapiro_analysis_1965}, it became clear that the coreness values were not normally distributed. Thus, to evaluate the session and the modality effect on the coreness and its associated properties, we fitted and tested an ANOVA using 5000 permutation-tests (lmPerm package in R).
Correlations between BCI scores and coreness were estimated via the use of repeated-measures correlations (rmcorr package in R \cite{bakdash_repeated_2017}). 

Results obtained from paired $t$-tests between conditions (to assess the condition effect) and from repeated-measures correlations referred to a statistical threshold of $0.05$ corrected for multiple comparisons by adopting a false discovery rate (FDR) criterion  \cite{benjamini_control_2001}, which is a method extensively used in biological studies \cite{mcauley_association_2009,sanders_toll-like_2012,matthews_using_2015}.

\section*{Results} 

\begin{figure*}
\begin{center}
\includegraphics[width=\textwidth]{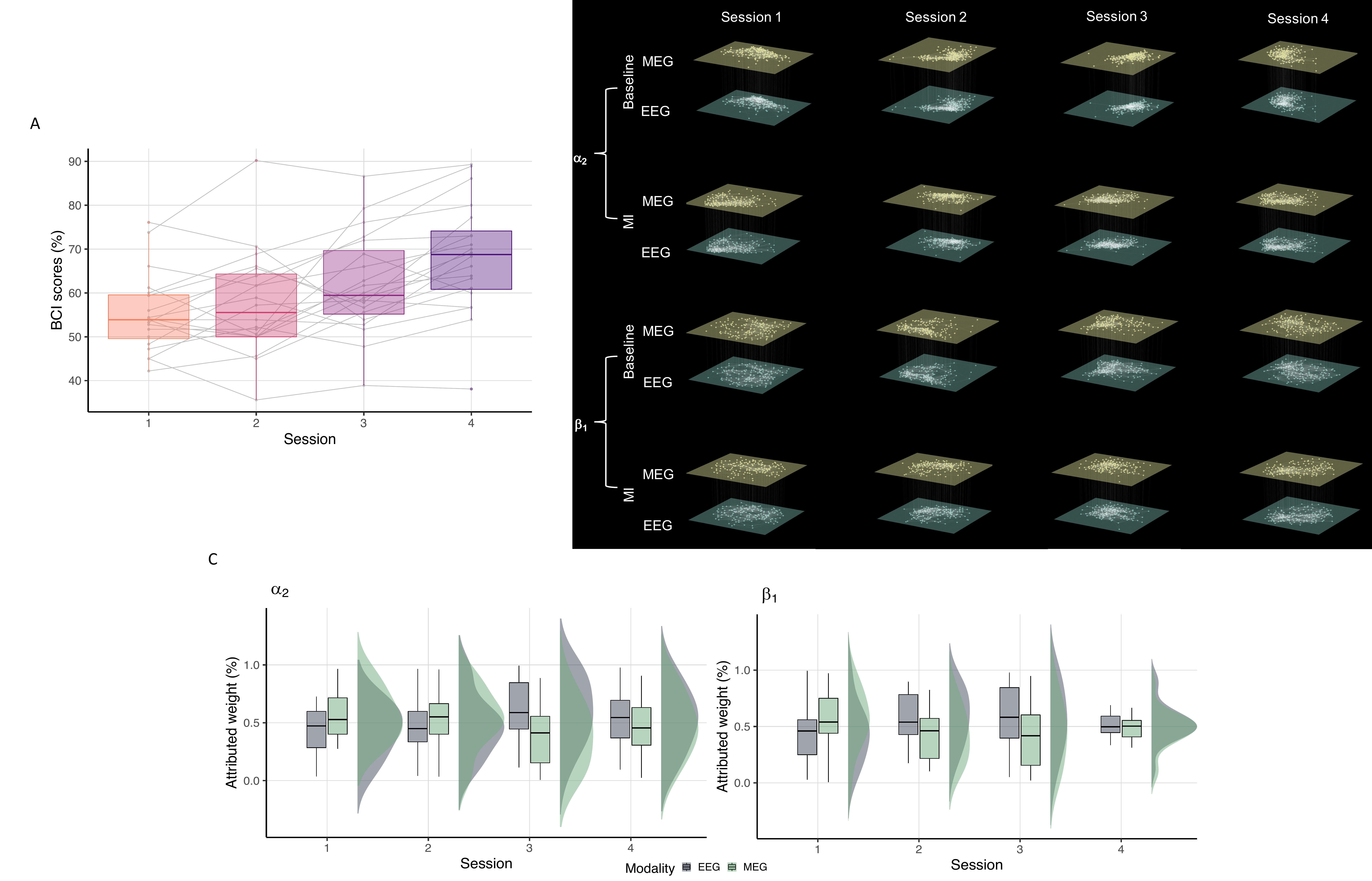}
	\caption{Behavioral performance and E/MEG contributions. (A) Distribution of BCI accuracy scores averaged across the runs of each session. Horizontal lines inside the box represent the median values. (B) Evolution of the E/MEG networks over sessions (average over the participants), obtained for each session, and condition within the $\alpha_2$ (top) and $\beta_1$ (bottom) ranges. (C) Evolution of attributed weights over sessions within the $\alpha_2$ (top) and $\beta_1$ (bottom) ranges. We plotted in grey and green the weight distribution associated, respectively, with EEG and MEG. Horizontal lines inside the box represent the median values.
	{\label{519840}}%
	}
\end{center}
\end{figure*}

Before studying the evolution of network properties over sessions, we first determined whether a learning effect was actually present. We applied a one-way repeated non-parametric ANOVA on the BCI accuracy scores averaged across the runs of each session with the session number as the intra-subject factor (Figure \ref{519840}A). Results confirmed that a learning effect was present at the group level (\(F(3,57)=13.9,\ p=6.56.10^{-7}\)). In particular, sixteen subjects out of 20 achieved  the ability to control the moving cursor by the end of the training, with accuracy scores above the chance level of $57 \%$~\cite{muller-putz_better_2008}. \\
As explained in the previous section, we started our analysis by using the adjacency matrices obtained from \cite{corsi_functional_2020} to build single layer networks (Figure {\ref{519840}B}, see Supplementary Materials \emph{Figures S1 \& S2}), and we investigated in which extent integrating the network properties obtained from EEG and MEG would be beneficial to the search of BCI training markers. \\
As a preliminary step, we studied the evolution of the attributed weights across sessions (Figure {\ref{519840}C}, see Supplementary Materials \emph{Figure S3}). We observed that the main session and modality effects occurred within the $\alpha$ and the $\beta$ bands, with significant interaction effects in $\alpha_2$ and $\beta_1$ bands (two-way ANOVA, respectively p=0.022 and p=0.027). In these bands, we observe similar trends. In session 1, $w_{MEG}$ is larger than $w_{EEG}$; then, the opposite effect occurred before the convergence to 0.5 at session 4. This final convergence to 0.5 indicates a progressive equal contribution of the two modality layers on the regional multiplex coreness.

\subsection*{Multiplex core-periphery provides additional information} 
\begin{figure*}
\begin{center}
		\includegraphics[width=\textwidth]{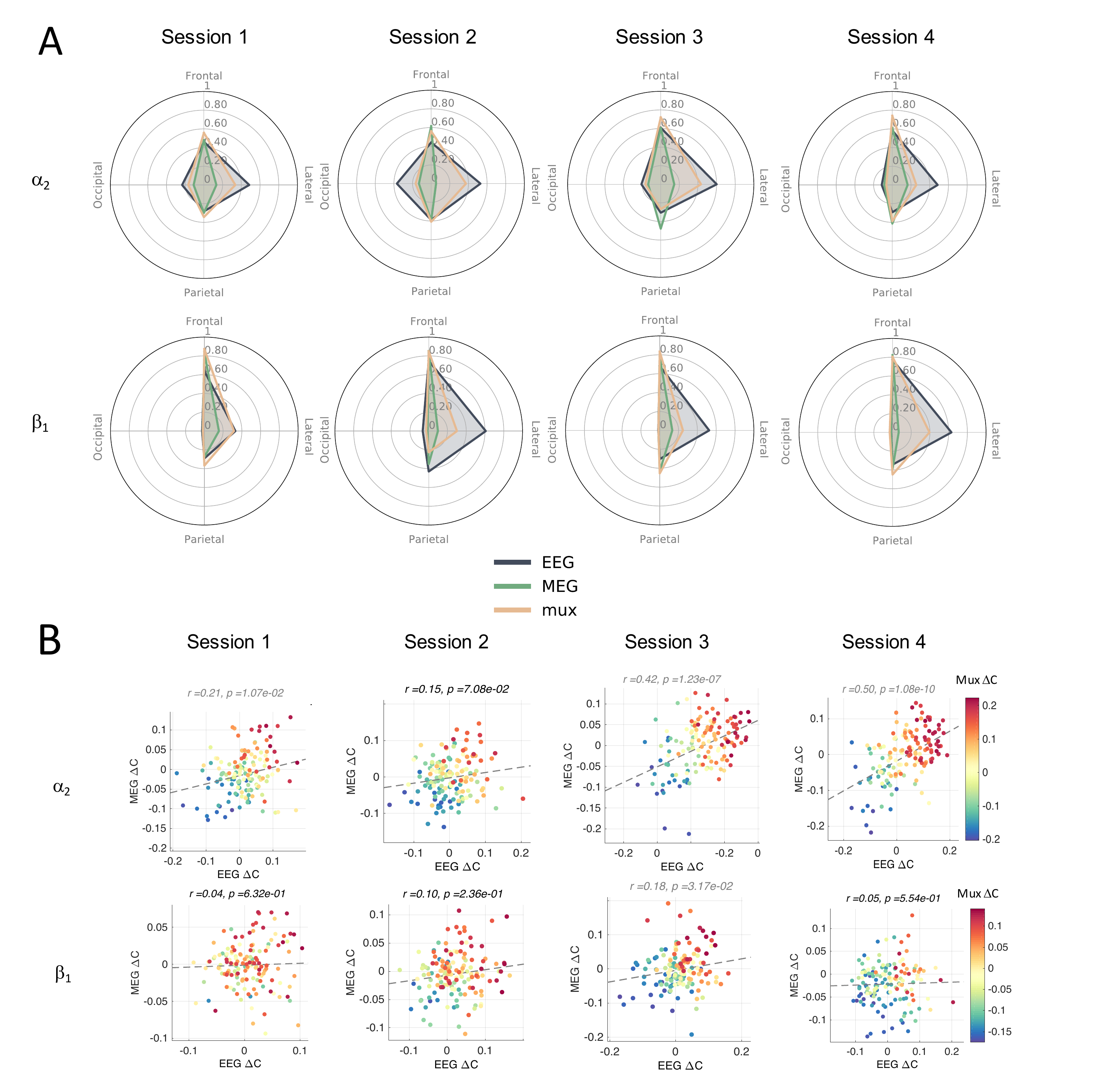}
		\caption{Single and multilayer provided information (A) Evolution of single layer and multiplex coreness values over BCI training in the MI condition. For a given axis associated with a single brain lobe, we plotted the median coreness value obtained across the subjects and the ROIs that belong to the lobe, respectively in EEG, MEG and multiplex (mux).  The first line corresponds to the evolution within the $\alpha_2$ frequency band and the second line corresponds to that within the $\beta_1$ frequency band.
		(B) Evolution of the relative coreness ($\Delta{C}$) over the sessions. On the X axis are represented the $\Delta{C}$ values, averaged over the subjects, obtained with the EEG layer; on the Y axis are presented the values obtained with the MEG layer. The color of the markers is associated with the values obtained with multiplex. Each marker corresponds to a given ROI.
		{\label{519843}}%
		}
\end{center}
\end{figure*}

We studied single and multiplex (mux) coreness trends over sessions in the MI condition (Figure {\ref{519843}}A). Similar tendencies were observed in the different modalities both within the $\alpha$ and $\beta$ frequency ranges (see Supplementary Materials \emph{Figure S4}). In particular, we observed that the highest values of MI coreness were obtained in ROIs that belongs to the frontal lobe. In $\alpha_2$, we obtained a progressive increase of the median value within the frontal lobe, especially in mux (see Figure {\ref{519843}}A). The second most important lobe was the lateral one, in particular for EEG and mux. We noticed an increase of the median value obtained within parietal lobe in MEG. In $\beta_1$, these observations were even clearer with an increase of the values obtained within the lateral lobes in EEG and mux, whereas values within in the parietal lobe were stable and those obtained within the occipital lobe were negligible.
These first observations showed that specific brain lobes presented clear variations of coreness values depending on the considered modality (for a more detailed presentation of the distribution of coreness values in the MI condition, see Supplementary Materials \emph{Figure S5}, and for a presentation of multiplex coreness values, see and Supplementary Materials \emph{Figure S6}).

The scatter plots represented in (Figure {\ref{519843}}B) are associated with the relative coreness ($\Delta{C}$) values obtained for each single layer (X and Y axis) and also for the multiplex. Within the $\alpha$ band, we observed that the distribution of points progressively followed a linear relationship between EEG and MEG $\Delta{C}$ values, meaning a non-negligible part of the information is shared by these modalities at the end of the training.
Within the $\beta$ band, we noticed an absence of a linear relationship between EEG and MEG, meaning that the two single layers shared less common information. 

Furthermore, we assessed the modality effect associated with $\Delta{C}$ via a one-way ANOVA, with the modality taken as the intra-subject factor. In the $\alpha - \beta$ ranges the parahippocampal gyrus significantly differed between modalities ($p < 0.030$) associated with visual functions \cite{aminoff_role_2013}. Within the $\alpha$ frequency range, we observed a significant modality effect in the middle-anterior part of the cingulate gyrus ($p < 0.030$) involved during decision making and memory consolidation \cite{kolling_multiple_2016}. In the $\beta_1$ band, the long insular gyrus, also associated with decision making \cite{uddin_structure_2017} presented a significant difference in terms of modality ($p < 0.001$).
The presented modality effects were driven by a significant difference between EEG and MEG relative coreness (Tukey $post-hoc$ multiple pairwise comparisons, $p$-values adjusted via the Holm method $p < 0.050$).

We also evaluated the information of interest provided by the multiplex with respect to single layers by statistically comparing the coreness of the MI versus the Rest conditions with a paired t-test ($p < 0.021$, see Supplementary Materials \emph{Tables S1-S3}).
We observed two opposite trends depending on the frequency range. 
In $\alpha_2$, at the single layer level, no consistent significant ROIs were obtained whereas we observed an increased involvement of the gyrus rectus with the multiplex with the training ($p < 0.01$ at session 4). This brain area is known to be associated with decision making involving a reward \cite{du_functional_2020}.
Within the $\beta$ ranges, we observed a lower number of ROIs showing a significant condition effect. In $\beta_1$, at the single layer, no significant ROIs ($p < 0.021$) were obtained during the first session whereas the multiplex presented three: short insular gyri (involved in motor plannning \cite{uddin_structure_2017}), planum polare of the superior temporal gyrus (deductive reasoning \cite{bonnefond_what_2012}), and the gyrus rectus (see Supplementary Materials \emph{Table S3}).

In the next sections, to directly account for the variations of coreness between conditions, we will focus our study on the relative coreness $\Delta{C}$. Furthermore, in order to take into account the most informative ROIs, we pre-selected the areas that show a significant condition effect at least once during the training before performing the analysis presented in the subsequent sections.

\subsection*{Relative coreness changes during training}  
To provide a more detailed description of the evolution of the relative coreness over training, we performed a one-way ANOVA for each layer separately (see Supplementary Materials \emph{Figures S8}).

We observed that $\Delta{C}$ presented a significant session effect involving different brain areas (Figure \ref{519849}A, see Supplementary Materials \emph{Figure S7}).
Within the $\alpha_2$ range, a significant session effect was observed in EEG mostly within the long insular gyrus and the gyrus rectus; a significant session effect was observed in MEG in the supramarginal gyrus (working memory and motor planning \cite{nee_meta-analysis_2013}); and in the multiplex a significant session effect was observed  in areas involved during motor planning and working memory (orbital part of the inferior frontal gyrus and subcallosal gyrus) \cite{milivojevic_functional_2009, wilson_orbitofrontal_2014,christophel_distributed_2017} and in learning complex motor skills (middle-posterior part of the cingulate gyrus)\cite{euston_role_2012}. 
In each case, we obtained an increase of $\Delta{C}$ with training (see Figure \ref{519849}A and Supplementary Materials \emph{Figure S7}). \\
Within the $\beta_1$ range, a significant session effect was observed in EEG within the inferior temporal gyrus (dual working memory task processing) and in the multiplex in areas associated with visual processing (superior temporal gyrus), working memory (middle frontal gyrus), and motor planning (short insular gyri). In the multiplex, most of the ROIs showing a significant session effect present a decrease of $\Delta{C}$ with training (see Figure \ref{519849}B).

\begin{figure*}
\begin{center}
		\includegraphics[width=\textwidth]{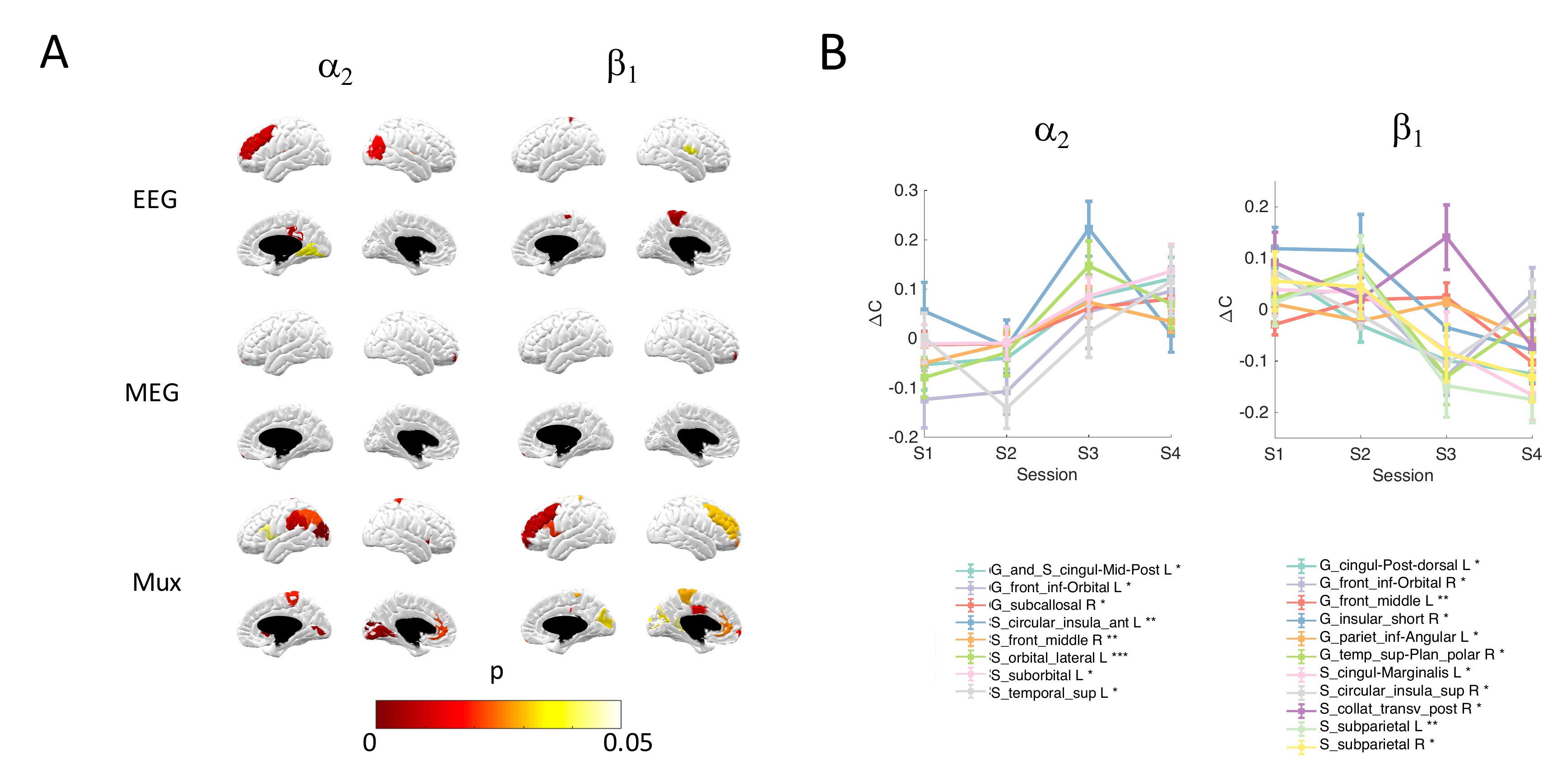}
		\caption{{Relative coreness changes during training. (A) ROIs showing a significant session effect (one-way ANOVA, $p < 0.05$). (B) Distribution over the training in the multiplex. Only the ROIs that present a significant session effect are represented (one-way ANOVA, $~p~<~0.05^{*}, ~p~<~0.01^{**}, ~p~<~0.001^{***}$).}
		{\label{519849}
		}}
\end{center}
\end{figure*}

\subsection*{Multiplex relative coreness correlated with future BCI performance} 
For the sake of simplicity, we will present our results only with relative coreness within the $\alpha_{2}$ band were the most significant observations were made. For a complete presentation of the results, see Supplementary Materials \emph{Figure S10-S11}.

We observed that the relative coreness presents a significant correlation with the BCI scores, within a larger number of significant ROIs in the multiplex in comparison with EEG or MEG (see Supplementary Materials \emph{Figure S10}). 
In EEG, negative correlations were obtained within the posterior-ventral part of the cingulate gyrus, the fronto-marginal gyrus ($p<0.01$) (respectively involved during learning a complex motor skill and working memory  \cite{euston_role_2012,stephan_functional_1995, johnson_selective_2002, solodkin_fine_2004}) and a positive correlation within the middle temporal gyrus (involved during the observation of motion \cite{rizzolatti_localization_1996}).
In MEG, a positive correlation was observed within the triangular part of the inferior frontal gyrus ($p<0.01$, involved during motor response inhibition and working memory \cite{milivojevic_functional_2009,wilson_orbitofrontal_2014, christophel_distributed_2017}) and a negative correlation within the cuneus (involved during visual processing \cite{silver_neural_2007}).
in the multiplex networks, positive correlations were obtained in regions involved respectively during motor tasks and motor imagery with working memory tasks (subcentral gyrus, superior parietal lobule, and subcallosal gyrus) \cite{yousry_localization_1997, solodkin_fine_2004,lotze_motor_2006, mcdougle_taking_2016}. A negative correlation was obtained within the gyrus rectus (decision making involving reward).

\begin{figure*}
\begin{center}
		\includegraphics[width=\textwidth]{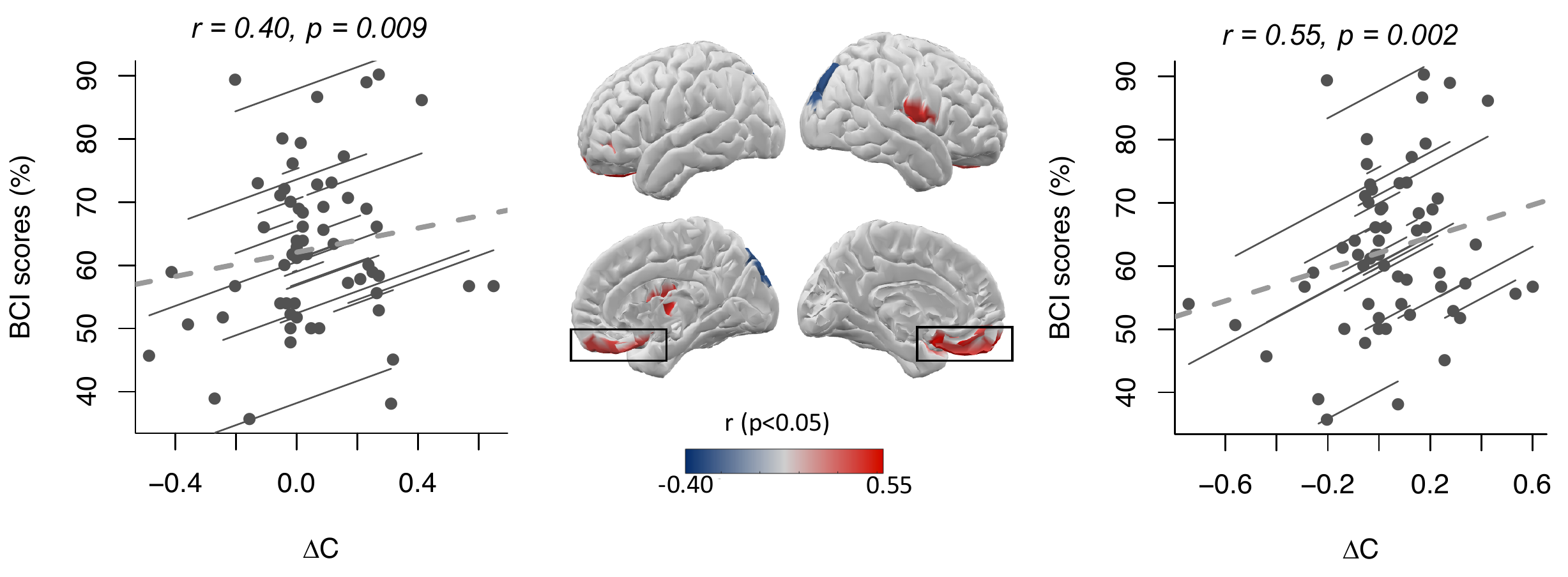}
		\caption{{Repeated correlations between BCI performance of the subsequent session and the multiplex relative coreness in the alpha2 band. At the center, we plotted the r-values projected onto the scalp ($p < 0.05$). On either side, scatter plots obtained from the two ROIs showing the highest r-values ($p < 0.01$). The dashed line represents the overall regression plot and the paralleled lines correspond to the fit to each subject’s data taken separately.
		{\label{519851}}%
		}}
\end{center}
\end{figure*}

To assess whether relative coreness could be associated with future BCI performance, we estimated the correlation between $\Delta{C}$ in session $i$ and the BCI score obtained in session $i+1$.
We observed significant correlations only with multiplex within the $\alpha_{2}$ band (Figure \ref{519851}). More precisely, a positive correlation ($p<0.01$) was observed in the gyrus rectus, the subcentral gyrus, but also the long insular gyrus (involved during somatosensory tasks \cite{uddin_structure_2017}). A negative correlation was obtained in the superior occipital gyrus associated with visual processing (in blue in Figure \ref{519851}) \cite{van_de_nieuwenhuijzen_meg-based_2013}.

\section*{Discussion}
Controlling a BCI remains a learned skill that is difficult to develop for a non-negligible number of users (15~\% - 30~\%) \cite{allison_could_2010}.
Previous studies dedicated to elicit neural dynamics underlying BCI skill acquisition in primates \cite{ganguly_emergence_2009,carmena_learning_2003} and humans  \cite{pichiorri_sensorimotor_2011,wander_distributed_2013} suggest the presence of a distributed and dynamic network of cortical areas above the motor-related ones. However, the evolution of such brain networks over training is largely unknown mainly because of a lack of longitudinal studies based on BCI paradigms \cite{perdikis_brain-machine_2020}.
Our protocol relied on reinforcement learning \cite{dayan_reinforcement_2008} based on a well-known two-target box task~\cite{wolpaw_wadsworth_2003} where a training effect has been obtained. In this study, we were particularly interested in understanding the brain network macroscale changes during the learning process. A few number of works, relying on BCI protocols and involving healthy subjects, have previously addressed this question \cite{corsi_functional_2020, stiso_learning_2020}. 

\subsection*{Tracking core-periphery changes}
It has been proved that core-periphery properties could be a valuable tool to track brain reorganization associated with cognitive processes \cite{van_den_heuvel_network_2013} but also disorders \cite{van_den_heuvel_exploring_2010,van_den_heuvel_abnormal_2013}. In this study, we worked with the coreness, a concise and robust metric that enables us to assess the likelihood to belong to the core of a network \cite{battiston_multilayer_2017,battiston_federico_multiplex_2018}. 
Regardless of the modality, opposite trends were obtained within the $\alpha$ and $\beta$ ranges in terms of the evolution of the discrimination between conditions and of the $\Delta{C}$ values with time (see Supplementary Materials \emph{Figure S8} and \emph{Tables S1-S3}). Nevertheless, these observations were particularly true for the multiplex involving $\alpha_2$ areas associated with somatosensory tasks and motor planning, and $\beta_1$ in areas associated with visual processing and working memory (see Figure\ref{519849}).

The $\alpha$ activity is known to be linked to the inhibition of task-irrelevant areas \cite{klimesch_eeg_2007, jensen_shaping_2010, haegens_-oscillations_2011}.
If $\beta$ desynchronization is clearly associated with sensorimotor tasks, recent studies suggest that $\beta$-synchrony maintains the current sensorimotor set \cite{spitzer_beyond_2017, engel_beta-band_2010}. In addition, $\beta$ activity is implicated in specific functions such as visual perception \cite{donner_population_2007, piantoni_beta_2010}
and working memory \cite{siegel_phase-dependent_2009},
and is associated with top-down controlled processing \cite{engel_beta-band_2010}.
From a functional connectivity perspective, in a previous work, we showed that MI-based BCI learning was associated with a progressive decrease of node strength in associative cortical regions and with the reinforcement of sensorimotor activity targeted by the experiment \cite{corsi_functional_2020}. In this case, $\alpha_2$ and $\beta_1$ shared a common behaviour.
Altogether, these results suggest a joint response of $\alpha_2$ with $\beta_1$ frequency bands during BCI training, associated with a reinforcement of the integration of sensorimotor areas in $\alpha_2$ paralleled with a functional connectivity release in the associative areas involved during visual processing and working memory in $\beta_1$.

\subsection*{Layer comparisons}
The complementary role of EEG and MEG has been proved at different levels: dipole orientation and source localization \cite{hamalainen_meg_theory_1993,sharon_advantage_2007, muthuraman_beamformer_2014} and subjects' mental state classification \cite{corsi_integrating_2018}. However, such complementarity has been poorly studied at the network level despite some interesting results in functional connectivity \cite{coquelet_comparing_2020}. 
To better capture network changes at different time or spatial scales, one can use multilayer models of networks \cite{de_domenico_mathematical_2013, boccaletti_structure_2014}. This approach enabled for example, in the time domain, to predict the relative learning rate via the flexibility \cite{bassett_dynamic_2011} in motor skill acquisition, but also to identify core-periphery changes in Alzheimer disease via a multimodal approach combining structural and functional networks \cite{battiston_multilayer_2017, battiston_federico_multiplex_2018, guillon_disrupted_2019}. \\
Here, based on previous work where MEG and DTI were combined \cite{guillon_disrupted_2019}, we integrated modalities knowing the contribution of each of them in such a way as to ensure the highest separation between conditions. These weights tended to converge to 0.5 (see Figure \ref{519840}C), meaning that the two modalities provided similar contribution to the multiplex network towards the end of the BCI training. This finding suggests that the two modalities are as important to discriminate MI and Rest conditions in the multiplex at the end of the training. As a result, the multiplex appeared to present a larger and more robust condition effect with respect to EEG and MEG (see Supplementary Materials \emph{Figures S7 and Tables S1-S3}). However, the attributed weights did not present a significant correlation with BCI performance. 
The approach proposed here also raised the possibility to compare results obtained from different layers. In particular, in $\alpha_2$, we obtained a progressive linear relationship between EEG and MEG relative corenesses with time over all the ROIs (see Figure \ref{519843}B).
This result suggests that, at a global level, MEG and EEG capture similar task-related processes occurring during the BCI experiment, especially at the end of training.
The modality effects, suggested in Figure \ref{519843}A, and actually observed at the relative coreness level, were driven by a significant difference between EEG and MEG relative corenesses. This effect was mostly observed in areas associated with decision making and memory consolidation, highlighting the utility to combine MEG and EEG networks to better capture mechanisms underlying learning process.

\subsection*{Markers of cognitive performance}
Identifying neural features underlying BCI performance is crucial to design optimized and individualized BCI architectures \cite{perdikis_subject-oriented_2014,de_vico_fallani_network_2019}. Among the elicited markers are psychological and demographical traits \cite{benaroch_are_2019}. From a neurophysiological perspective, previous studies identified power spectra in $\theta$, $\alpha$ and $\gamma$ bands as potential predictors of BCI scores \cite{ahn_high_2013, jeunet_predicting_2015}. In our study, the most significant results were obtained in $\alpha_{2}$ and $\beta_{1}$ frequency bands.
Recent findings proved that functional connectivity could correlate with the user's performance ~\cite{sugata_alpha_2014,pichiorri_brain-computer_2015,de_vico_fallani_multiscale_2013}. However, these studies were associated with a single session BCI performance.
In a recent work, we showed that the regional connectivity strength of specific associative cortical areas could explain the BCI performance in the same session but also the future learning rate \cite{corsi_functional_2020}. Here, we were particularly interested in identifying markers of BCI performance at the core-periphery network level. 

If EEG, MEG, and multiplex presented associations with BCI scores, only the latter presents a significant correlation with the BCI performance of the next session based on the relative coreness within the $\alpha_{2}$ band (see Supplementary Materials \emph{Figure S10-S11}). Two trends were again observed: a positive correlation in areas respectively involved during decision making and somatosensory tasks (gyrus rectus, subcentral gyrus, and long insular gyrus) and a negative correlation in the superior occipital gyrus associated with visual processing (see Figure \ref{519851}). These findings are in line with previous studies that reported a larger clustering coefficient in the gyrus rectus associated with a higher nodal betweenness centrality (NBC) in sensorimotor areas and a reduced NBC in visual areas in the context of motor training \cite{li_probabilistic_2014, calmels_neural_2020}. Altogether, these results support the hypothesis that sensorimotor areas and associative areas play a crucial role in motor sequence learning as well as in abstract task learning~\cite{mcdougle_taking_2016,hetu_neural_2013,hardwick_neural_2018,dayan_neuroplasticity_2011} and that cognitive processes involved in the supervisory attentional system \cite{van_zomeren_clinical_1994,wolpert_motor_2012} are important to perform MI tasks \cite{guillot_neurophysiological_2010} and  motor learning \cite{wulf_attention_2007, lohse_role_2014, dayan_learning_2000,gottlieb_attention_2012}.

\subsection*{Caveats and perspectives}
The temporal window of study is a crucial matter when considering a longitudinal experimentation, especially in the BCI domain. Our participants followed a four-session-training program, within two weeks. This temporal window might not be sufficient to observe the full learning process~\cite{yin_dynamic_2009, perdikis_brain-machine_2020}. However our results constitute the first observations of a training process at the core-periphery level. Further studies based on longer BCI training are necessary to assess whether our observations could be still verified on a larger temporal scale.

This work could pave the way to further explore of the integration of M/EEG network information to better understand neural mechanisms underlying learning but also task performance in particular in the use of BCI in a clinical context. However, before considering multimodal BCIs in routine, further developments are required to increase MEG portability. The use of new generation of MEG sensors (i.e. optically-pumped magnetometers) could meet this need \cite{boto_measurements_2017,labyt_magnetoencephalography_2018, boto_wearable_2019, tierney_optically_2019}.

\section*{Conclusion}
In this work, we have proved that studying the network integration changes at the single and multilayer levels provides additional information to  characterize dynamic brain reorganization during BCI training. We found that a progressive increase of the integration of somatosensory areas in the $\alpha$ band was paralleled by a decrease of the integration of visual processing and working memory areas in the $\beta$ band. 
More importantly, these changes were more visible in multiplex in which brain network properties correlated with future BCI scores in the $\alpha_2$ band.
Taken together, our results cast new light on brain network reorganization occurring during BCI training and more generally during human learning.

%\pagebreak
%\clearpage

%\section*{References}
%\textbf{copy paste .bib file here}
\bibliographystyle{nature_mag}
\bibliography{Manuscript_revised.bib}

\begin{thebibliography}{100}
\expandafter\ifx\csname url\endcsname\relax
  \def\url#1{\texttt{#1}}\fi
\expandafter\ifx\csname urlprefix\endcsname\relax\def\urlprefix{URL }\fi
\providecommand{\bibinfo}[2]{#2}
\providecommand{\eprint}[2][]{\url{#2}}

\bibitem{reddy_brain_2018}
\bibinfo{author}{Reddy, P.~G.} \emph{et~al.}
\newblock \bibinfo{title}{Brain state flexibility accompanies motor-skill
  acquisition}.
\newblock \emph{\bibinfo{journal}{Neuroimage}} \textbf{\bibinfo{volume}{171}},
  \bibinfo{pages}{135--147} (\bibinfo{year}{2018}).
\newblock
  \urlprefix\url{http://www.sciencedirect.com/science/article/pii/S1053811917311175}.

\bibitem{deng_resting-state_2016}
\bibinfo{author}{Deng, Z.}, \bibinfo{author}{Chandrasekaran, B.},
  \bibinfo{author}{Wang, S.} \& \bibinfo{author}{Wong, P.~C.}
\newblock \bibinfo{title}{Resting-state low-frequency fluctuations reflect
  individual differences in spoken language learning}.
\newblock \emph{\bibinfo{journal}{Cortex}} \textbf{\bibinfo{volume}{76}},
  \bibinfo{pages}{63--78} (\bibinfo{year}{2016}).
\newblock
  \urlprefix\url{https://www.ncbi.nlm.nih.gov/pmc/articles/PMC4777637/}.

\bibitem{sheppard_large-scale_2012}
\bibinfo{author}{Sheppard, J.~P.}, \bibinfo{author}{Wang, J.-P.} \&
  \bibinfo{author}{Wong, P. C.~M.}
\newblock \bibinfo{title}{Large-scale {Cortical} {Network} {Properties}
  {Predict} {Future} {Sound}-to-{Word} {Learning} {Success}}.
\newblock \emph{\bibinfo{journal}{Journal of Cognitive Neuroscience}}
  \textbf{\bibinfo{volume}{24}}, \bibinfo{pages}{1087--1103}
  (\bibinfo{year}{2012}).
\newblock
  \urlprefix\url{https://www.ncbi.nlm.nih.gov/pmc/articles/PMC3736731/}.

\bibitem{sami_graph_2013}
\bibinfo{author}{Sami, S.} \& \bibinfo{author}{Miall, R.~C.}
\newblock \bibinfo{title}{Graph network analysis of immediate motor-learning
  induced changes in resting state {BOLD}}.
\newblock \emph{\bibinfo{journal}{Frontiers in Human Neuroscience}}
  \textbf{\bibinfo{volume}{7}} (\bibinfo{year}{2013}).
\newblock
  \urlprefix\url{https://www.ncbi.nlm.nih.gov/pmc/articles/PMC3654214/}.

\bibitem{taubert_long-term_2011}
\bibinfo{author}{Taubert, M.}, \bibinfo{author}{Lohmann, G.},
  \bibinfo{author}{Margulies, D.~S.}, \bibinfo{author}{Villringer, A.} \&
  \bibinfo{author}{Ragert, P.}
\newblock \bibinfo{title}{Long-term effects of motor training on resting-state
  networks and underlying brain structure}.
\newblock \emph{\bibinfo{journal}{Neuroimage}} \textbf{\bibinfo{volume}{57}},
  \bibinfo{pages}{1492--1498} (\bibinfo{year}{2011}).
\newblock \urlprefix\url{https://doi.org/10.1016/j.neuroimage.2011.05.078}.

\bibitem{ge_motor_2015}
\bibinfo{author}{Ge, R.}, \bibinfo{author}{Zhang, H.}, \bibinfo{author}{Yao,
  L.} \& \bibinfo{author}{Long, Z.}
\newblock \bibinfo{title}{Motor imagery learning induced changes in functional
  connectivity of the default mode network}.
\newblock \emph{\bibinfo{journal}{IEEE Transactions on Neural Systems and
  Rehabilitation Engineering}} \textbf{\bibinfo{volume}{23}},
  \bibinfo{pages}{138--148} (\bibinfo{year}{2015}).
\newblock \urlprefix\url{https://ieeexplore.ieee.org/document/6849455}.

\bibitem{mcdougle_taking_2016}
\bibinfo{author}{McDougle, S.~D.}, \bibinfo{author}{Ivry, R.~B.} \&
  \bibinfo{author}{Taylor, J.~A.}
\newblock \bibinfo{title}{Taking {Aim} at the {Cognitive} {Side} of {Learning}
  in {Sensorimotor} {Adaptation} {Tasks}}.
\newblock \emph{\bibinfo{journal}{Trends in Cognitive Sciences}}
  \textbf{\bibinfo{volume}{20}}, \bibinfo{pages}{535--544}
  (\bibinfo{year}{2016}).
\newblock \urlprefix\url{http://dx.doi.org/10.1016/j.tics.2016.05.002}.

\bibitem{rizzolatti_cortical_2001}
\bibinfo{author}{Rizzolatti, G.} \& \bibinfo{author}{Luppino, G.}
\newblock \bibinfo{title}{The cortical motor system}.
\newblock \emph{\bibinfo{journal}{Neuron}} \textbf{\bibinfo{volume}{31}},
  \bibinfo{pages}{889--901} (\bibinfo{year}{2001}).
\newblock \urlprefix\url{http://dx.doi.org/10.1016/s0896-6273(01)00423-8}.

\bibitem{katiuscia_reorganization_2009}
\bibinfo{author}{Katiuscia, S.} \emph{et~al.}
\newblock \bibinfo{title}{Reorganization and enhanced functional connectivity
  of motor areas in repetitive ankle movements after training in locomotor
  attention}.
\newblock \emph{\bibinfo{journal}{Brain Research}}
  \textbf{\bibinfo{volume}{1297}}, \bibinfo{pages}{124--134}
  (\bibinfo{year}{2009}).
\newblock
  \urlprefix\url{http://www.sciencedirect.com/science/article/pii/S0006899309017661}.

\bibitem{heitger_motor_2012}
\bibinfo{author}{Heitger, M.~H.} \emph{et~al.}
\newblock \bibinfo{title}{Motor learning-induced changes in functional brain
  connectivity as revealed by means of graph-theoretical network analysis}.
\newblock \emph{\bibinfo{journal}{Neuroimage}} \textbf{\bibinfo{volume}{61}},
  \bibinfo{pages}{633--650} (\bibinfo{year}{2012}).
\newblock \urlprefix\url{https://doi.org/10.1016%2Fj.neuroimage.2012.03.067}.

\bibitem{sporns_modular_2016}
\bibinfo{author}{Sporns, O.} \& \bibinfo{author}{Betzel, R.~F.}
\newblock \bibinfo{title}{Modular {Brain} {Networks}}.
\newblock \emph{\bibinfo{journal}{Annual Review of Psychology}}
  \textbf{\bibinfo{volume}{67}}, \bibinfo{pages}{613--640}
  (\bibinfo{year}{2016}).
\newblock
  \urlprefix\url{http://dx.doi.org/10.1146/annurev-psych-122414-033634}.

\bibitem{gallen_brain_2019}
\bibinfo{author}{Gallen, C.~L.} \& \bibinfo{author}{D'Esposito, M.}
\newblock \bibinfo{title}{Brain {Modularity}: {A} {Biomarker} of
  {Intervention}-related {Plasticity}}.
\newblock \emph{\bibinfo{journal}{Trends in Cognitive Sciences}}
  (\bibinfo{year}{2019}).
\newblock
  \urlprefix\url{http://www.sciencedirect.com/science/article/pii/S1364661319300427}.

\bibitem{bassett_dynamic_2011}
\bibinfo{author}{Bassett, D.~S.} \emph{et~al.}
\newblock \bibinfo{title}{Dynamic reconfiguration of human brain networks
  during learning}.
\newblock \emph{\bibinfo{journal}{Proceedings of the National Academy of
  Sciences of the United States of America}} \textbf{\bibinfo{volume}{108}},
  \bibinfo{pages}{7641--7646} (\bibinfo{year}{2011}).
\newblock \urlprefix\url{http://www.Proceedings of the National Academy of
  Sciences of the United States of America.org/content/108/18/7641}.

\bibitem{bassett_learning-induced_2015}
\bibinfo{author}{Bassett, D.~S.}, \bibinfo{author}{Yang, M.},
  \bibinfo{author}{Wymbs, N.~F.} \& \bibinfo{author}{Grafton, S.~T.}
\newblock \bibinfo{title}{Learning-induced autonomy of sensorimotor systems}.
\newblock \emph{\bibinfo{journal}{Nature Neuroscience}}
  \textbf{\bibinfo{volume}{18}}, \bibinfo{pages}{744--751}
  (\bibinfo{year}{2015}).
\newblock \urlprefix\url{http://dx.doi.org/10.1038/nn.3993}.

\bibitem{hiremath_brain_2015}
\bibinfo{author}{Hiremath, S.~V.} \emph{et~al.}
\newblock \bibinfo{title}{Brain computer interface learning for systems based
  on electrocorticography and intracortical microelectrode arrays}.
\newblock \emph{\bibinfo{journal}{Frontiers in Integrative Neuroscience}}
  \bibinfo{pages}{40} (\bibinfo{year}{2015}).
\newblock \urlprefix\url{http://dx.doi.org/10.3389/fnint.2015.00040}.

\bibitem{sitaram_closed-loop_2016}
\bibinfo{author}{Sitaram, R.} \emph{et~al.}
\newblock \bibinfo{title}{Closed-loop brain training: the science of
  neurofeedback}.
\newblock \emph{\bibinfo{journal}{Nature Reviews Neuroscience}}
  \textbf{\bibinfo{volume}{18}}, \bibinfo{pages}{86--100}
  (\bibinfo{year}{2016}).
\newblock \urlprefix\url{https://doi.org/10.1038%2Fnrn.2016.164}.

\bibitem{ito_self-reorganization_2020}
\bibinfo{author}{Ito, H.}, \bibinfo{author}{Fujiki, S.}, \bibinfo{author}{Mori,
  Y.} \& \bibinfo{author}{Kansaku, K.}
\newblock \bibinfo{title}{Self-reorganization of neuronal activation patterns
  in the cortex under brain-machine interface and neural operant conditioning}.
\newblock \emph{\bibinfo{journal}{Neuroscience Research}}
  (\bibinfo{year}{2020}).
\newblock \urlprefix\url{http://dx.doi.org/10.1016/j.neures.2020.03.008}.

\bibitem{orsborn_parsing_2017}
\bibinfo{author}{Orsborn, A.~L.} \& \bibinfo{author}{Pesaran, B.}
\newblock \bibinfo{title}{Parsing learning in networks using brain-machine
  interfaces}.
\newblock \emph{\bibinfo{journal}{Current Opinion in Neurobiology}}
  \textbf{\bibinfo{volume}{46}}, \bibinfo{pages}{76--83}
  (\bibinfo{year}{2017}).
\newblock \urlprefix\url{http://dx.doi.org/10.1016/j.conb.2017.08.002}.

\bibitem{wander_distributed_2013}
\bibinfo{author}{Wander, J.~D.} \emph{et~al.}
\newblock \bibinfo{title}{Distributed cortical adaptation during learning of a
  brain-computer interface task}.
\newblock \emph{\bibinfo{journal}{Proceedings of the National Academy of
  Sciences of the United States of America}} \textbf{\bibinfo{volume}{110}},
  \bibinfo{pages}{10818--10823} (\bibinfo{year}{2013}).
\newblock \urlprefix\url{https://doi.org/10.1073%2FProceedings of the National
  Academy of Sciences of the United States of America.1221127110}.

\bibitem{pichiorri_sensorimotor_2011}
\bibinfo{author}{Pichiorri, F.} \emph{et~al.}
\newblock \bibinfo{title}{Sensorimotor rhythm-based brain--computer interface
  training: the impact on motor cortical responsiveness}.
\newblock \emph{\bibinfo{journal}{Journal of Neural Engineering}}
  \textbf{\bibinfo{volume}{8}}, \bibinfo{pages}{025020} (\bibinfo{year}{2011}).
\newblock \urlprefix\url{http://dx.doi.org/10.1088/1741-2560/8/2/025020}.

\bibitem{borgatti_models_2000}
\bibinfo{author}{Borgatti, S.~P.} \& \bibinfo{author}{Everett, M.~G.}
\newblock \bibinfo{title}{Models of core/periphery structures}.
\newblock \emph{\bibinfo{journal}{Social Networks}}
  \textbf{\bibinfo{volume}{21}}, \bibinfo{pages}{375--395}
  (\bibinfo{year}{2000}).
\newblock
  \urlprefix\url{http://www.sciencedirect.com/science/article/pii/S0378873399000192}.

\bibitem{girvan_community_2002}
\bibinfo{author}{Girvan, M.} \& \bibinfo{author}{Newman, M. E.~J.}
\newblock \bibinfo{title}{Community structure in social and biological
  networks}.
\newblock \emph{\bibinfo{journal}{Proceedings of the National Academy of
  Sciences of the United States of America}} \textbf{\bibinfo{volume}{99}},
  \bibinfo{pages}{7821--7826} (\bibinfo{year}{2002}).
\newblock \urlprefix\url{https://www.Proceedings of the National Academy of
  Sciences of the United States of America.org/content/99/12/7821}.
\newblock \bibinfo{note}{Publisher: National Academy of Sciences Section:
  Physical Sciences}.

\bibitem{de_domenico_mathematical_2013}
\bibinfo{author}{De~Domenico, M.} \emph{et~al.}
\newblock \bibinfo{title}{Mathematical {Formulation} of {Multilayer}
  {Networks}}.
\newblock \emph{\bibinfo{journal}{Physical Review X}}
  \textbf{\bibinfo{volume}{3}}, \bibinfo{pages}{041022} (\bibinfo{year}{2013}).
\newblock \urlprefix\url{https://link.aps.org/doi/10.1103/PhysRevX.3.041022}.

\bibitem{boccaletti_structure_2014}
\bibinfo{author}{Boccaletti, S.} \emph{et~al.}
\newblock \bibinfo{title}{The structure and dynamics of multilayer networks}.
\newblock \emph{\bibinfo{journal}{Physics Reports}}
  \textbf{\bibinfo{volume}{544}}, \bibinfo{pages}{1--122}
  (\bibinfo{year}{2014}).
\newblock
  \urlprefix\url{http://www.sciencedirect.com/science/article/pii/S0370157314002105}.

\bibitem{battiston_multilayer_2017}
\bibinfo{author}{Battiston, F.}, \bibinfo{author}{Nicosia, V.},
  \bibinfo{author}{Chavez, M.} \& \bibinfo{author}{Latora, V.}
\newblock \bibinfo{title}{Multilayer motif analysis of brain networks}.
\newblock \emph{\bibinfo{journal}{Chaos: An Interdisciplinary Journal of
  Nonlinear Science}} \textbf{\bibinfo{volume}{27}}, \bibinfo{pages}{047404}
  (\bibinfo{year}{2017}).
\newblock \urlprefix\url{https://aip.scitation.org/doi/10.1063/1.4979282}.

\bibitem{battiston_federico_multiplex_2018}
\bibinfo{author}{{Battiston Federico}}, \bibinfo{author}{{Guillon Jeremy}},
  \bibinfo{author}{{Chavez Mario}}, \bibinfo{author}{{Latora Vito}} \&
  \bibinfo{author}{{De Vico Fallani Fabrizio}}.
\newblock \bibinfo{title}{Multiplex core--periphery organization of the human
  connectome}.
\newblock \emph{\bibinfo{journal}{Journal of The Royal Society Interface}}
  \textbf{\bibinfo{volume}{15}}, \bibinfo{pages}{20180514}
  (\bibinfo{year}{2018}).
\newblock
  \urlprefix\url{https://royalsocietypublishing.org/doi/full/10.1098/rsif.2018.0514}.

\bibitem{de_domenico_mapping_2016}
\bibinfo{author}{De~Domenico, M.}, \bibinfo{author}{Sasai, S.} \&
  \bibinfo{author}{Arenas, A.}
\newblock \bibinfo{title}{Mapping {Multiplex} {Hubs} in {Human} {Functional}
  {Brain} {Networks}}.
\newblock \emph{\bibinfo{journal}{Frontiers in Neuroscience}}
  \textbf{\bibinfo{volume}{10}} (\bibinfo{year}{2016}).
\newblock
  \urlprefix\url{https://www.ncbi.nlm.nih.gov/pmc/articles/PMC4945645/}.

\bibitem{de_domenico_multilayer_2017}
\bibinfo{author}{De~Domenico, M.}
\newblock \bibinfo{title}{Multilayer modeling and analysis of human brain
  networks}.
\newblock \emph{\bibinfo{journal}{Gigascience}} \textbf{\bibinfo{volume}{6}},
  \bibinfo{pages}{1--8} (\bibinfo{year}{2017}).
\newblock
  \urlprefix\url{https://www.ncbi.nlm.nih.gov/pmc/articles/PMC5437946/}.

\bibitem{tewarie_integrating_2016}
\bibinfo{author}{Tewarie, P.} \emph{et~al.}
\newblock \bibinfo{title}{Integrating cross-frequency and within band
  functional networks in resting-state {MEG}: {A} multi-layer network
  approach}.
\newblock \emph{\bibinfo{journal}{Neuroimage}} \textbf{\bibinfo{volume}{142}},
  \bibinfo{pages}{324--336} (\bibinfo{year}{2016}).
\newblock \urlprefix\url{http://dx.doi.org/10.1016/j.neuroimage.2016.07.057}.

\bibitem{guillon_loss_2017}
\bibinfo{author}{Guillon, J.} \emph{et~al.}
\newblock \bibinfo{title}{Loss of brain inter-frequency hubs in {Alzheimer}'s
  disease}.
\newblock \emph{\bibinfo{journal}{Scientific Reports}}
  \textbf{\bibinfo{volume}{7}} (\bibinfo{year}{2017}).
\newblock \urlprefix\url{https://www.nature.com/articles/s41598-017-07846-w}.

\bibitem{buldu_frequency-based_2018}
\bibinfo{author}{Buld{\'u}, J.~M.} \& \bibinfo{author}{Porter, M.~A.}
\newblock \bibinfo{title}{Frequency-based brain networks: {From} a multiplex
  framework to a full multilayer description}.
\newblock \emph{\bibinfo{journal}{Network Neuroscience}}
  \textbf{\bibinfo{volume}{2}}, \bibinfo{pages}{418--441}
  (\bibinfo{year}{2018}).
\newblock
  \urlprefix\url{https://www.ncbi.nlm.nih.gov/pmc/articles/PMC6147638/}.

\bibitem{guillon_disrupted_2019}
\bibinfo{author}{Guillon, J.} \emph{et~al.}
\newblock \bibinfo{title}{Disrupted core-periphery structure of multimodal
  brain networks in {Alzheimer}'s disease}.
\newblock \emph{\bibinfo{journal}{Network Neuroscience}}
  \textbf{\bibinfo{volume}{3}}, \bibinfo{pages}{635--652}
  (\bibinfo{year}{2019}).
\newblock \urlprefix\url{http://dx.doi.org/10.1162/netn_a_00087}.
\newblock \bibinfo{note}{Publisher: MIT Press}.

\bibitem{cuffin_comparison_1979}
\bibinfo{author}{Cuffin, B.~N.} \& \bibinfo{author}{Cohen, D.}
\newblock \bibinfo{title}{Comparison of the magnetoencephalogram and
  electroencephalogram}.
\newblock \emph{\bibinfo{journal}{Electroencephalography and Clinical
  Neurophysiology}} \textbf{\bibinfo{volume}{47}}, \bibinfo{pages}{132--146}
  (\bibinfo{year}{1979}).

\bibitem{geisler_surface_1961}
\bibinfo{author}{Geisler, C.~D.} \& \bibinfo{author}{Gerstein, G.~L.}
\newblock \bibinfo{title}{The surface {EEG} in relation to its sources}.
\newblock \emph{\bibinfo{journal}{Electroencephalography and Clinical
  Neurophysiology}} \textbf{\bibinfo{volume}{13}}, \bibinfo{pages}{927--934}
  (\bibinfo{year}{1961}).
\newblock \urlprefix\url{http://dx.doi.org/10.1016/0013-4694(61)90199-7}.

\bibitem{delucchi_scalp_1962}
\bibinfo{author}{Delucchi, M.~R.}, \bibinfo{author}{Garoutte, B.} \&
  \bibinfo{author}{Aird, R.~B.}
\newblock \bibinfo{title}{The scalp as an electroencephalographic averager}.
\newblock \emph{\bibinfo{journal}{Electroencephalography and Clinical
  Neurophysiology}} \textbf{\bibinfo{volume}{14}}, \bibinfo{pages}{191--196}
  (\bibinfo{year}{1962}).

\bibitem{hamalainen_meg_theory_1993}
\bibinfo{author}{Hämäläinen, M.}, \bibinfo{author}{Hari, R.},
  \bibinfo{author}{Ilmoniemi, R.~J.}, \bibinfo{author}{Knuutila, J.} \&
  \bibinfo{author}{Lounasmaa, O.~V.}
\newblock \bibinfo{title}{Magnetoencephalography-theory, instrumentation, and
  applications to noninvasive studies of the working human brain}.
\newblock \emph{\bibinfo{journal}{Reviews of Modern Physics}}
  \textbf{\bibinfo{volume}{65}}, \bibinfo{pages}{413--497}
  (\bibinfo{year}{1993}).
\newblock \urlprefix\url{http://dx.doi.org/10.1103/RevModPhys.65.413}.

\bibitem{wood_electrical_1985}
\bibinfo{author}{Wood, C.~C.}, \bibinfo{author}{Cohen, D.},
  \bibinfo{author}{Cuffin, B.~N.}, \bibinfo{author}{Yarita, M.} \&
  \bibinfo{author}{Allison, T.}
\newblock \bibinfo{title}{Electrical sources in human somatosensory cortex:
  identification by combined magnetic and potential recordings}.
\newblock \emph{\bibinfo{journal}{Science (New York, N.Y.)}}
  \textbf{\bibinfo{volume}{227}}, \bibinfo{pages}{1051--1053}
  (\bibinfo{year}{1985}).
\newblock
  \urlprefix\url{https://science.sciencemag.org/content/227/4690/1051.long}.

\bibitem{sharon_advantage_2007}
\bibinfo{author}{Sharon, D.}, \bibinfo{author}{H{\"a}m{\"a}l{\"a}inen, M.~S.},
  \bibinfo{author}{Tootell, R. B.~H.}, \bibinfo{author}{Halgren, E.} \&
  \bibinfo{author}{Belliveau, J.~W.}
\newblock \bibinfo{title}{The advantage of combining {MEG} and {EEG}:
  {Comparison} to {fMRI} in focally stimulated visual cortex}.
\newblock \emph{\bibinfo{journal}{Neuroimage}} \textbf{\bibinfo{volume}{36}},
  \bibinfo{pages}{1225--1235} (\bibinfo{year}{2007}).
\newblock \urlprefix\url{http://dx.doi.org/10.1016/j.neuroimage.2007.03.066}.

\bibitem{corsi_integrating_2018}
\bibinfo{author}{Corsi, M.-C.} \emph{et~al.}
\newblock \bibinfo{title}{Integrating {EEG} and {MEG} {Signals} to {Improve}
  {Motor} {Imagery} {Classification} in {Brain}–{Computer} {Interface}}.
\newblock \emph{\bibinfo{journal}{International Journal of Neural Systems}}
  \textbf{\bibinfo{volume}{29}}, \bibinfo{pages}{1850014}
  (\bibinfo{year}{2018}).
\newblock
  \urlprefix\url{https://www.worldscientific.com/doi/abs/10.1142/S0129065718500144}.

\bibitem{wolpaw_wadsworth_2003}
\bibinfo{author}{Wolpaw, J.~R.}, \bibinfo{author}{McFarland, D.~J.},
  \bibinfo{author}{Vaughan, T.~M.} \& \bibinfo{author}{Schalk, G.}
\newblock \bibinfo{title}{The {Wadsworth} {Center} brain-computer interface
  ({BCI}) research and development program}.
\newblock \emph{\bibinfo{journal}{IEEE Transactions on Neural Systems and
  Rehabilitation Engineering}} \textbf{\bibinfo{volume}{11}},
  \bibinfo{pages}{204--207} (\bibinfo{year}{2003}).
\newblock \urlprefix\url{http://dx.doi.org/10.1109/TNSRE.2003.814442}.

\bibitem{schalk_bci2000:_2004}
\bibinfo{author}{Schalk, G.}, \bibinfo{author}{McFarland, D.~J.},
  \bibinfo{author}{Hinterberger, T.}, \bibinfo{author}{Birbaumer, N.} \&
  \bibinfo{author}{Wolpaw, J.~R.}
\newblock \bibinfo{title}{{BCI}2000: a general-purpose brain-computer interface
  ({BCI}) system}.
\newblock \emph{\bibinfo{journal}{IEEE Transactions on Biomedical Engineering}}
  \textbf{\bibinfo{volume}{51}}, \bibinfo{pages}{1034--1043}
  (\bibinfo{year}{2004}).
\newblock \urlprefix\url{http://dx.doi.org/10.1109/TBME.2004.827072}.

\bibitem{oostenveld_fieldtrip:_2010}
\bibinfo{author}{Oostenveld, R.}, \bibinfo{author}{Fries, P.},
  \bibinfo{author}{Maris, E.}, \bibinfo{author}{Schoffelen, J.-M.} \&
  \bibinfo{author}{Oostenveld}.
\newblock \bibinfo{title}{{FieldTrip}: {Open} {Source} {Software} for
  {Advanced} {Analysis} of {MEG}, {EEG}, and {Invasive} {Electrophysiological}
  {Data}}.
\newblock \emph{\bibinfo{journal}{Computational Intelligence and Neuroscience}}
  \textbf{\bibinfo{volume}{2011}}, \bibinfo{pages}{e156869}
  (\bibinfo{year}{2010}).
\newblock \urlprefix\url{http://www.hindawi.com/journals/cin/2011/156869/abs/}.

\bibitem{gross_good_2013}
\bibinfo{author}{Gross, J.} \emph{et~al.}
\newblock \bibinfo{title}{Good practice for conducting and reporting {MEG}
  research}.
\newblock \emph{\bibinfo{journal}{Neuroimage}} \textbf{\bibinfo{volume}{65}},
  \bibinfo{pages}{349--363} (\bibinfo{year}{2013}).
\newblock \urlprefix\url{http://dx.doi.org/10.1016/j.neuroimage.2012.10.001}.

\bibitem{fischl_freesurfer_2012}
\bibinfo{author}{Fischl, B.}
\newblock \bibinfo{title}{{FreeSurfer}}.
\newblock \emph{\bibinfo{journal}{Neuroimage}} \textbf{\bibinfo{volume}{62}},
  \bibinfo{pages}{774--781} (\bibinfo{year}{2012}).
\newblock
  \urlprefix\url{http://www.sciencedirect.com/science/article/pii/S1053811912000389}.

\bibitem{tadel_brainstorm:_2011}
\bibinfo{author}{Tadel, F.}, \bibinfo{author}{Baillet, S.},
  \bibinfo{author}{Mosher, J.~C.}, \bibinfo{author}{Pantazis, D.} \&
  \bibinfo{author}{Leahy, R.~M.}
\newblock \bibinfo{title}{Brainstorm: {A} {User}-{Firendly} {Application} for
  {MEG}/{EEG} {Analysis}}.
\newblock \emph{\bibinfo{journal}{Computational Intelligence and Neuroscience}}
  \textbf{\bibinfo{volume}{2011}} (\bibinfo{year}{2011}).
\newblock \urlprefix\url{http://dx.doi.org/10.1155/2011/879716}.

\bibitem{corsi_functional_2020}
\bibinfo{author}{Corsi, M.-C.} \emph{et~al.}
\newblock \bibinfo{title}{Functional disconnection of associative cortical
  areas predicts performance during {BCI} training}.
\newblock \emph{\bibinfo{journal}{Neuroimage}} \textbf{\bibinfo{volume}{209}},
  \bibinfo{pages}{116500} (\bibinfo{year}{2020}).
\newblock
  \urlprefix\url{http://www.sciencedirect.com/science/article/pii/S1053811919310912}.

\bibitem{taulu_spatiotemporal_2006}
\bibinfo{author}{Taulu, S.} \& \bibinfo{author}{Simola, J.}
\newblock \bibinfo{title}{Spatiotemporal signal space separation method for
  rejecting nearby interference in {MEG} measurements}.
\newblock \emph{\bibinfo{journal}{Physics in Medicine \& Biology}}
  \textbf{\bibinfo{volume}{51}}, \bibinfo{pages}{1759--1768}
  (\bibinfo{year}{2006}).
\newblock \urlprefix\url{http://dx.doi.org/10.1088/0031-9155/51/7/008}.

\bibitem{bell_information-maximization_1995}
\bibinfo{author}{Bell, A.~J.} \& \bibinfo{author}{Sejnowski, T.~J.}
\newblock \bibinfo{title}{An information-maximization approach to blind
  separation and blind deconvolution}.
\newblock \emph{\bibinfo{journal}{Neural Computation}}
  \textbf{\bibinfo{volume}{7}}, \bibinfo{pages}{1129--1159}
  (\bibinfo{year}{1995}).

\bibitem{fuchs_boundary_2001}
\bibinfo{author}{Fuchs, M.}, \bibinfo{author}{Wagner, M.} \&
  \bibinfo{author}{Kastner, J.}
\newblock \bibinfo{title}{Boundary element method volume conductor models for
  {EEG} source reconstruction}.
\newblock \emph{\bibinfo{journal}{Clinical Neurophysiology}}
  \textbf{\bibinfo{volume}{112}}, \bibinfo{pages}{1400--1407}
  (\bibinfo{year}{2001}).
\newblock
  \urlprefix\url{http://www.sciencedirect.com/science/article/pii/S1388245701005892}.

\bibitem{gramfort_openmeeg:_2010}
\bibinfo{author}{Gramfort, A.}, \bibinfo{author}{Papadopoulo, T.},
  \bibinfo{author}{Olivi, E.} \& \bibinfo{author}{Clerc, M.}
\newblock \bibinfo{title}{{OpenMEEG}: opensource software for quasistatic
  bioelectromagnetics}.
\newblock \emph{\bibinfo{journal}{BioMedical Engineering OnLine}}
  \textbf{\bibinfo{volume}{9}}, \bibinfo{pages}{45} (\bibinfo{year}{2010}).
\newblock \urlprefix\url{http://dx.doi.org/10.1186/1475-925X-9-45}.

\bibitem{fuchs_linear_1999}
\bibinfo{author}{Fuchs, M.}, \bibinfo{author}{Wagner, M.},
  \bibinfo{author}{K{\"o}hler, T.} \& \bibinfo{author}{Wischmann, H.-A.}
\newblock \bibinfo{title}{Linear and nonlinear current density
  reconstructions}.
\newblock \emph{\bibinfo{journal}{Journal of Clinical Neurophysiology}}
  \textbf{\bibinfo{volume}{16}}, \bibinfo{pages}{267--295}
  (\bibinfo{year}{1999}).

\bibitem{lin_assessing_2006}
\bibinfo{author}{Lin, F.-H.} \emph{et~al.}
\newblock \bibinfo{title}{Assessing and improving the spatial accuracy in {MEG}
  source localization by depth-weighted minimum-norm estimates}.
\newblock \emph{\bibinfo{journal}{Neuroimage}} \textbf{\bibinfo{volume}{31}},
  \bibinfo{pages}{160--171} (\bibinfo{year}{2006}).
\newblock \urlprefix\url{http://dx.doi.org/10.1016/j.neuroimage.2005.11.054}.

\bibitem{gramfort_mne_2014}
\bibinfo{author}{Gramfort, A.} \emph{et~al.}
\newblock \bibinfo{title}{{MNE} software for processing {MEG} and {EEG} data}.
\newblock \emph{\bibinfo{journal}{Neuroimage}} \textbf{\bibinfo{volume}{86}},
  \bibinfo{pages}{446--460} (\bibinfo{year}{2014}).
\newblock \urlprefix\url{http://dx.doi.org/10.1016/j.neuroimage.2013.10.027}.

\bibitem{destrieux_automatic_2010}
\bibinfo{author}{Destrieux, C.}, \bibinfo{author}{Fischl, B.},
  \bibinfo{author}{Dale, A.} \& \bibinfo{author}{Halgren, E.}
\newblock \bibinfo{title}{Automatic parcellation of human cortical gyri and
  sulci using standard anatomical nomenclature}.
\newblock \emph{\bibinfo{journal}{Neuroimage}} \textbf{\bibinfo{volume}{53}},
  \bibinfo{pages}{1--15} (\bibinfo{year}{2010}).
\newblock
  \urlprefix\url{https://www.ncbi.nlm.nih.gov/pmc/articles/PMC2937159/}.

\bibitem{sekihara_removal_2011}
\bibinfo{author}{Sekihara, K.}, \bibinfo{author}{Owen, J.},
  \bibinfo{author}{Trisno, S.} \& \bibinfo{author}{Nagarajan, S.~S.}
\newblock \bibinfo{title}{Removal of spurious coherence in {MEG} source-space
  coherence analysis}.
\newblock \emph{\bibinfo{journal}{IEEE transactions on bio-medical
  engineering}} \textbf{\bibinfo{volume}{58}}, \bibinfo{pages}{3121--3129}
  (\bibinfo{year}{2011}).
\newblock
  \urlprefix\url{https://www.ncbi.nlm.nih.gov/pmc/articles/PMC4096348/}.

\bibitem{kennedy_particle_1995}
\bibinfo{author}{Kennedy, J.} \& \bibinfo{author}{Eberhart, R.}
\newblock \bibinfo{title}{Particle swarm optimization}.
\newblock In \emph{\bibinfo{booktitle}{Proceedings of {ICNN}'95 -
  {International} {Conference} on {Neural} {Networks}}},
  vol.~\bibinfo{volume}{4}, \bibinfo{pages}{1942--1948 vol.4}
  (\bibinfo{year}{1995}).
\newblock \urlprefix\url{http://dx.doi.org/10.1109/ICNN.1995.488968}.
\newblock \bibinfo{note}{ISSN: null}.

\bibitem{rubinov_complex_2010}
\bibinfo{author}{Rubinov, M.} \& \bibinfo{author}{Sporns, O.}
\newblock \bibinfo{title}{Complex network measures of brain connectivity:
  {Uses} and interpretations}.
\newblock \emph{\bibinfo{journal}{Neuroimage}} \textbf{\bibinfo{volume}{52}},
  \bibinfo{pages}{1059--1069} (\bibinfo{year}{2010}).
\newblock
  \urlprefix\url{http://www.sciencedirect.com/science/article/pii/S105381190901074X}.

\bibitem{klimesch_eeg_1999}
\bibinfo{author}{Klimesch, W.}
\newblock \bibinfo{title}{{EEG} alpha and theta oscillations reflect cognitive
  and memory performance: a review and analysis}.
\newblock \emph{\bibinfo{journal}{Brain Research Reviews}}
  \textbf{\bibinfo{volume}{29}}, \bibinfo{pages}{169--195}
  (\bibinfo{year}{1999}).
\newblock
  \urlprefix\url{http://www.sciencedirect.com/science/article/pii/S0165017398000563}.

\bibitem{pichiorri_brain-computer_2015}
\bibinfo{author}{Pichiorri, F.} \emph{et~al.}
\newblock \bibinfo{title}{Brain-computer interface boosts motor imagery
  practice during stroke recovery}.
\newblock \emph{\bibinfo{journal}{Annals of Neurology}}
  \textbf{\bibinfo{volume}{77}}, \bibinfo{pages}{851--865}
  (\bibinfo{year}{2015}).
\newblock \urlprefix\url{http://dx.doi.org/10.1002/ana.24390}.

\bibitem{shapiro_analysis_1965}
\bibinfo{author}{Shapiro, S.~S.} \& \bibinfo{author}{Wilk, M.~B.}
\newblock \bibinfo{title}{An analysis of variance test for normality (complete
  samples)}.
\newblock \emph{\bibinfo{journal}{Biometrika}} \textbf{\bibinfo{volume}{52}},
  \bibinfo{pages}{591--611} (\bibinfo{year}{1965}).
\newblock
  \urlprefix\url{https://academic.oup.com/biomet/article/52/3-4/591/336553}.
\newblock \bibinfo{note}{Publisher: Oxford Academic}.

\bibitem{bakdash_repeated_2017}
\bibinfo{author}{Bakdash, J.~Z.} \& \bibinfo{author}{Marusich, L.~R.}
\newblock \bibinfo{title}{Repeated {Measures} {Correlation}}.
\newblock \emph{\bibinfo{journal}{Frontiers in Psychology}}
  \textbf{\bibinfo{volume}{8}} (\bibinfo{year}{2017}).
\newblock
  \urlprefix\url{https://www.frontiersin.org/articles/10.3389/fpsyg.2017.00456/full}.

\bibitem{benjamini_control_2001}
\bibinfo{author}{Benjamini, Y.} \& \bibinfo{author}{Yekutieli, D.}
\newblock \bibinfo{title}{The {Control} of the {False} {Discovery} {Rate} in
  {Multiple} {Testing} under {Dependency}}.
\newblock \emph{\bibinfo{journal}{The Annals of Statistics}}
  \textbf{\bibinfo{volume}{29}}, \bibinfo{pages}{1165--1188}
  (\bibinfo{year}{2001}).
\newblock \urlprefix\url{https://www.jstor.org/stable/2674075}.

\bibitem{mcauley_association_2009}
\bibinfo{author}{McAuley, E.~Z.} \emph{et~al.}
\newblock \bibinfo{title}{Association between the serotonin 2a receptor gene
  and bipolar affective disorder in an {Australian} cohort}.
\newblock \emph{\bibinfo{journal}{Psychiatric Genetics}}
  \textbf{\bibinfo{volume}{19}}, \bibinfo{pages}{244--252}
  (\bibinfo{year}{2009}).

\bibitem{sanders_toll-like_2012}
\bibinfo{author}{Sanders, M.~S.}, \bibinfo{author}{van Well, G.~T.},
  \bibinfo{author}{Ouburg, S.}, \bibinfo{author}{Morr\'e, S.~A.} \&
  \bibinfo{author}{van Furth, A.~M.}
\newblock \bibinfo{title}{Toll-like receptor 9polymorphisms are associated with
  severity variables in a cohort of meningococcal meningitis survivors}.
\newblock \emph{\bibinfo{journal}{BMC Infectious Diseases}}
  \textbf{\bibinfo{volume}{12}}, \bibinfo{pages}{112} (\bibinfo{year}{2012}).
\newblock \urlprefix\url{https://doi.org/10.1186/1471-2334-12-112}.

\bibitem{matthews_using_2015}
\bibinfo{author}{Matthews, D.~E.} \& \bibinfo{author}{Farewell, V.~T.}
\newblock \emph{\bibinfo{title}{Using and {Understanding} {Medical}
  {Statistics}}} (\bibinfo{publisher}{Karger Medical and Scientific
  Publishers}, \bibinfo{year}{2015}).
\newblock \bibinfo{note}{Google-Books-ID: DL8nCgAAQBAJ}.

\bibitem{muller-putz_better_2008}
\bibinfo{author}{M{\"u}ller-putz, G.~R.}, \bibinfo{author}{Scherer, R.},
  \bibinfo{author}{Brunner, C.}, \bibinfo{author}{Leeb, R.} \&
  \bibinfo{author}{Pfurtscheller, G.}
\newblock \bibinfo{title}{Better than random: a closer look on {BCI} results}.
\newblock \emph{\bibinfo{journal}{International Journal of
  Bioelectromagnetism}} \textbf{\bibinfo{volume}{10}} (\bibinfo{year}{2008}).
\newblock
  \urlprefix\url{http://citeseerx.ist.psu.edu/viewdoc/download?doi=10.1.1.330.3349&rep=rep1&type=pdf}.

\bibitem{aminoff_role_2013}
\bibinfo{author}{Aminoff, E.~M.}, \bibinfo{author}{Kveraga, K.} \&
  \bibinfo{author}{Bar, M.}
\newblock \bibinfo{title}{The role of the parahippocampal cortex in cognition}.
\newblock \emph{\bibinfo{journal}{Trends in Cognitive Sciences}}
  \textbf{\bibinfo{volume}{17}}, \bibinfo{pages}{379--390}
  (\bibinfo{year}{2013}).
\newblock
  \urlprefix\url{https://www.ncbi.nlm.nih.gov/pmc/articles/PMC3786097/}.

\bibitem{kolling_multiple_2016}
\bibinfo{author}{Kolling, N.}, \bibinfo{author}{Behrens, T.~E.},
  \bibinfo{author}{Wittmann, M.~K.} \& \bibinfo{author}{Rushworth, M.~F.}
\newblock \bibinfo{title}{Multiple signals in anterior cingulate cortex}.
\newblock \emph{\bibinfo{journal}{Current Opinion in Neurobiology}}
  \textbf{\bibinfo{volume}{37}}, \bibinfo{pages}{36--43}
  (\bibinfo{year}{2016}).
\newblock
  \urlprefix\url{http://www.sciencedirect.com/science/article/pii/S0959438815001853}.

\bibitem{uddin_structure_2017}
\bibinfo{author}{Uddin, L.~Q.}, \bibinfo{author}{Nomi, J.~S.},
  \bibinfo{author}{Hebert-Seropian, B.}, \bibinfo{author}{Ghaziri, J.} \&
  \bibinfo{author}{Boucher, O.}
\newblock \bibinfo{title}{Structure and function of the human insula}.
\newblock \emph{\bibinfo{journal}{Journal of Clinical Neurophysiology}}
  \textbf{\bibinfo{volume}{34}}, \bibinfo{pages}{300--306}
  (\bibinfo{year}{2017}).
\newblock
  \urlprefix\url{https://www.ncbi.nlm.nih.gov/pmc/articles/PMC6032992/}.

\bibitem{du_functional_2020}
\bibinfo{author}{Du, J.} \emph{et~al.}
\newblock \bibinfo{title}{Functional connectivity of the orbitofrontal cortex,
  anterior cingulate cortex, and inferior frontal gyrus in humans}.
\newblock \emph{\bibinfo{journal}{Cortex}} \textbf{\bibinfo{volume}{123}},
  \bibinfo{pages}{185--199} (\bibinfo{year}{2020}).
\newblock
  \urlprefix\url{http://www.sciencedirect.com/science/article/pii/S001094521930365X}.

\bibitem{bonnefond_what_2012}
\bibinfo{author}{Bonnefond, M.} \emph{et~al.}
\newblock \bibinfo{title}{What {MEG} can reveal about inference making: {The}
  case of if...then sentences}.
\newblock \emph{\bibinfo{journal}{Human Brain Mapping}}
  \textbf{\bibinfo{volume}{34}}, \bibinfo{pages}{684--697}
  (\bibinfo{year}{2012}).
\newblock
  \urlprefix\url{https://www.ncbi.nlm.nih.gov/pmc/articles/PMC6870271/}.

\bibitem{nee_meta-analysis_2013}
\bibinfo{author}{Nee, D.~E.} \emph{et~al.}
\newblock \bibinfo{title}{A {Meta}-analysis of {Executive} {Components} of
  {Working} {Memory}}.
\newblock \emph{\bibinfo{journal}{Cerebral Cortex}}
  \textbf{\bibinfo{volume}{23}}, \bibinfo{pages}{264--282}
  (\bibinfo{year}{2013}).
\newblock
  \urlprefix\url{https://academic.oup.com/cercor/article/23/2/264/283011}.
\newblock \bibinfo{note}{Publisher: Oxford Academic}.

\bibitem{milivojevic_functional_2009}
\bibinfo{author}{Milivojevic, B.}, \bibinfo{author}{Hamm, J.~P.} \&
  \bibinfo{author}{Corballis, M.~C.}
\newblock \bibinfo{title}{Functional neuroanatomy of mental rotation}.
\newblock \emph{\bibinfo{journal}{Journal of Cognitive Neuroscience}}
  \textbf{\bibinfo{volume}{21}}, \bibinfo{pages}{945--959}
  (\bibinfo{year}{2009}).
\newblock \urlprefix\url{http://dx.doi.org/10.1162/jocn.2009.21085}.

\bibitem{wilson_orbitofrontal_2014}
\bibinfo{author}{Wilson, R.~C.}, \bibinfo{author}{Takahashi, Y.~K.},
  \bibinfo{author}{Schoenbaum, G.} \& \bibinfo{author}{Niv, Y.}
\newblock \bibinfo{title}{Orbitofrontal {Cortex} as a {Cognitive} {Map} of
  {Task} {Space}}.
\newblock \emph{\bibinfo{journal}{Neuron}} \textbf{\bibinfo{volume}{81}},
  \bibinfo{pages}{267--279} (\bibinfo{year}{2014}).
\newblock
  \urlprefix\url{http://www.sciencedirect.com/science/article/pii/S0896627313010398}.

\bibitem{christophel_distributed_2017}
\bibinfo{author}{Christophel, T.~B.}, \bibinfo{author}{Klink, P.~C.},
  \bibinfo{author}{Spitzer, B.}, \bibinfo{author}{Roelfsema, P.~R.} \&
  \bibinfo{author}{Haynes, J.-D.}
\newblock \bibinfo{title}{The {Distributed} {Nature} of {Working} {Memory}}.
\newblock \emph{\bibinfo{journal}{Trends in Cognitive Sciences}}
  \textbf{\bibinfo{volume}{21}}, \bibinfo{pages}{111--124}
  (\bibinfo{year}{2017}).
\newblock
  \urlprefix\url{http://www.sciencedirect.com/science/article/pii/S1364661316302170}.

\bibitem{euston_role_2012}
\bibinfo{author}{Euston, D.~R.}, \bibinfo{author}{Gruber, A.~J.} \&
  \bibinfo{author}{McNaughton, B.~L.}
\newblock \bibinfo{title}{The {Role} of {Medial} {Prefrontal} {Cortex} in
  {Memory} and {Decision} {Making}}.
\newblock \emph{\bibinfo{journal}{Neuron}} \textbf{\bibinfo{volume}{76}},
  \bibinfo{pages}{1057--1070} (\bibinfo{year}{2012}).
\newblock
  \urlprefix\url{https://www.cell.com/neuron/abstract/S0896-6273(12)01108-7}.

\bibitem{stephan_functional_1995}
\bibinfo{author}{Stephan, K.~M.} \emph{et~al.}
\newblock \bibinfo{title}{Functional anatomy of the mental representation of
  upper extremity movements in healthy subjects}.
\newblock \emph{\bibinfo{journal}{Journal of Neurophysiology}}
  \textbf{\bibinfo{volume}{73}}, \bibinfo{pages}{373--386}
  (\bibinfo{year}{1995}).
\newblock \urlprefix\url{http://dx.doi.org/10.1152/jn.1995.73.1.373}.

\bibitem{johnson_selective_2002}
\bibinfo{author}{Johnson, S.~H.} \emph{et~al.}
\newblock \bibinfo{title}{Selective activation of a parietofrontal circuit
  during implicitly imagined prehension}.
\newblock \emph{\bibinfo{journal}{Neuroimage}} \textbf{\bibinfo{volume}{17}},
  \bibinfo{pages}{1693--1704} (\bibinfo{year}{2002}).
\newblock
  \urlprefix\url{https://www.sciencedirect.com/science/article/abs/pii/S1053811902912656?via\%3Dihub}.

\bibitem{solodkin_fine_2004}
\bibinfo{author}{Solodkin, A.}, \bibinfo{author}{Hlustik, P.},
  \bibinfo{author}{Chen, E.~E.} \& \bibinfo{author}{Small, S.~L.}
\newblock \bibinfo{title}{Fine modulation in network activation during motor
  execution and motor imagery}.
\newblock \emph{\bibinfo{journal}{Cerebral Cortex}}
  \textbf{\bibinfo{volume}{14}}, \bibinfo{pages}{1246--1255}
  (\bibinfo{year}{2004}).
\newblock \urlprefix\url{http://dx.doi.org/10.1093/cercor/bhh086}.

\bibitem{rizzolatti_localization_1996}
\bibinfo{author}{Rizzolatti, G.} \emph{et~al.}
\newblock \bibinfo{title}{Localization of grasp representations in humans by
  {PET}: 1. {Observation} versus execution}.
\newblock \emph{\bibinfo{journal}{Experimental Brain Research}}
  \textbf{\bibinfo{volume}{111}}, \bibinfo{pages}{246--252}
  (\bibinfo{year}{1996}).
\newblock \urlprefix\url{https://doi.org/10.1007/BF00227301}.

\bibitem{silver_neural_2007}
\bibinfo{author}{Silver, M.~A.}, \bibinfo{author}{Ress, D.} \&
  \bibinfo{author}{Heeger, D.~J.}
\newblock \bibinfo{title}{Neural correlates of sustained spatial attention in
  human early visual cortex}.
\newblock \emph{\bibinfo{journal}{Journal of Neurophysiology}}
  \textbf{\bibinfo{volume}{97}}, \bibinfo{pages}{229--237}
  (\bibinfo{year}{2007}).
\newblock \urlprefix\url{http://dx.doi.org/10.1152/jn.00677.2006}.

\bibitem{yousry_localization_1997}
\bibinfo{author}{Yousry, T.~A.} \emph{et~al.}
\newblock \bibinfo{title}{Localization of the motor hand area to a knob on the
  precentral gyrus. {A} new landmark.}
\newblock \emph{\bibinfo{journal}{Brain}} \textbf{\bibinfo{volume}{120}},
  \bibinfo{pages}{141--157} (\bibinfo{year}{1997}).
\newblock
  \urlprefix\url{https://academic.oup.com/brain/article/120/1/141/312820}.

\bibitem{lotze_motor_2006}
\bibinfo{author}{Lotze, M.} \& \bibinfo{author}{Halsband, U.}
\newblock \bibinfo{title}{Motor imagery}.
\newblock \emph{\bibinfo{journal}{Journal of Physiology-Paris}}
  \textbf{\bibinfo{volume}{99}}, \bibinfo{pages}{386--395}
  (\bibinfo{year}{2006}).
\newblock
  \urlprefix\url{http://www.sciencedirect.com/science/article/pii/S0928425706000210}.

\bibitem{van_de_nieuwenhuijzen_meg-based_2013}
\bibinfo{author}{van~de Nieuwenhuijzen, M.~E.} \emph{et~al.}
\newblock \bibinfo{title}{{MEG}-based decoding of the spatiotemporal dynamics
  of visual category perception}.
\newblock \emph{\bibinfo{journal}{Neuroimage}} \textbf{\bibinfo{volume}{83}},
  \bibinfo{pages}{1063--1073} (\bibinfo{year}{2013}).
\newblock
  \urlprefix\url{https://www.sciencedirect.com/science/article/abs/pii/S1053811913008446?via\%3Dihub}.

\bibitem{allison_could_2010}
\bibinfo{author}{Allison, B.~Z.} \& \bibinfo{author}{Neuper, C.}
\newblock \bibinfo{title}{Could {Anyone} {Use} a {BCI}?}
\newblock In \bibinfo{editor}{Tan, D.~S.} \& \bibinfo{editor}{Nijholt, A.}
  (eds.) \emph{\bibinfo{booktitle}{Brain-{Computer} {Interfaces}}},
  Human-{Computer} {Interaction} {Series}, \bibinfo{pages}{35--54}
  (\bibinfo{publisher}{Springer London}, \bibinfo{year}{2010}).
\newblock
  \urlprefix\url{http://link.springer.com/chapter/10.1007/978-1-84996-272-8_3}.

\bibitem{ganguly_emergence_2009}
\bibinfo{author}{Ganguly, K.} \& \bibinfo{author}{Carmena, J.~M.}
\newblock \bibinfo{title}{Emergence of a stable cortical map for
  neuroprosthetic control.}
\newblock \emph{\bibinfo{journal}{PLoS Biology}} \textbf{\bibinfo{volume}{7}},
  \bibinfo{pages}{e1000153} (\bibinfo{year}{2009}).
\newblock
  \urlprefix\url{http://journals.plos.org/plosbiology/article?id=10.1371/journal.pbio.1000153}.

\bibitem{carmena_learning_2003}
\bibinfo{author}{Carmena, J.~M.} \emph{et~al.}
\newblock \bibinfo{title}{Learning to control a brain-machine interface for
  reaching and grasping by primates.}
\newblock \emph{\bibinfo{journal}{PLoS Biology}} \textbf{\bibinfo{volume}{1}},
  \bibinfo{pages}{E42} (\bibinfo{year}{2003}).
\newblock
  \urlprefix\url{http://journals.plos.org/plosbiology/article?id=10.1371/journal.pbio.0000042}.

\bibitem{perdikis_brain-machine_2020}
\bibinfo{author}{Perdikis, S.} \& \bibinfo{author}{Millan, J. d.~R.}
\newblock \bibinfo{title}{Brain-{Machine} {Interfaces}: {A} {Tale} of {Two}
  {Learners}}.
\newblock \emph{\bibinfo{journal}{IEEE Systems, Man, and Cybernetics Magazine}}
  \textbf{\bibinfo{volume}{6}}, \bibinfo{pages}{12--19} (\bibinfo{year}{2020}).
\newblock \urlprefix\url{https://ieeexplore.ieee.org/document/9141465}.
\newblock \bibinfo{note}{IEEE Systems, Man, and Cybernetics Magazine}.

\bibitem{dayan_reinforcement_2008}
\bibinfo{author}{Dayan, P.} \& \bibinfo{author}{Niv, Y.}
\newblock \bibinfo{title}{Reinforcement learning: the good, the bad and the
  ugly}.
\newblock \emph{\bibinfo{journal}{Current Opinion in Neurobiology}}
  \textbf{\bibinfo{volume}{18}}, \bibinfo{pages}{185--196}
  (\bibinfo{year}{2008}).
\newblock \urlprefix\url{http://dx.doi.org/10.1016/j.conb.2008.08.003}.

\bibitem{stiso_learning_2020}
\bibinfo{author}{Stiso, J.} \emph{et~al.}
\newblock \bibinfo{title}{Learning in brain-computer interface control
  evidenced by joint decomposition of brain and behavior}.
\newblock \emph{\bibinfo{journal}{Journal of Neural Engineering}}
  (\bibinfo{year}{2020}).
\newblock \urlprefix\url{http://dx.doi.org/10.1088/1741-2552/ab9064}.

\bibitem{van_den_heuvel_network_2013}
\bibinfo{author}{van~den Heuvel, M.~P.} \& \bibinfo{author}{Sporns, O.}
\newblock \bibinfo{title}{Network hubs in the human brain}.
\newblock \emph{\bibinfo{journal}{Trends in Cognitive Sciences}}
  \textbf{\bibinfo{volume}{17}}, \bibinfo{pages}{683--696}
  (\bibinfo{year}{2013}).
\newblock \urlprefix\url{http://dx.doi.org/10.1016/j.tics.2013.09.012}.

\bibitem{van_den_heuvel_exploring_2010}
\bibinfo{author}{van~den Heuvel, M.~P.} \& \bibinfo{author}{Hulshoff~Pol,
  H.~E.}
\newblock \bibinfo{title}{Exploring the brain network: {A} review on
  resting-state {fMRI} functional connectivity}.
\newblock \emph{\bibinfo{journal}{European Neuropsychopharmacology}}
  \textbf{\bibinfo{volume}{20}}, \bibinfo{pages}{519--534}
  (\bibinfo{year}{2010}).
\newblock
  \urlprefix\url{http://www.sciencedirect.com/science/article/pii/S0924977X10000684}.

\bibitem{van_den_heuvel_abnormal_2013}
\bibinfo{author}{van~den Heuvel, M.~P.} \emph{et~al.}
\newblock \bibinfo{title}{Abnormal rich club organization and functional brain
  dynamics in schizophrenia}.
\newblock \emph{\bibinfo{journal}{JAMA Psychiatry}}
  \textbf{\bibinfo{volume}{70}}, \bibinfo{pages}{783--792}
  (\bibinfo{year}{2013}).
\newblock \urlprefix\url{http://dx.doi.org/10.1001/jamapsychiatry.2013.1328}.

\bibitem{klimesch_eeg_2007}
\bibinfo{author}{Klimesch, W.}, \bibinfo{author}{Sauseng, P.} \&
  \bibinfo{author}{Hanslmayr, S.}
\newblock \bibinfo{title}{{EEG} alpha oscillations: the inhibition-timing
  hypothesis}.
\newblock \emph{\bibinfo{journal}{Brain Res Rev}}
  \textbf{\bibinfo{volume}{53}}, \bibinfo{pages}{63--88}
  (\bibinfo{year}{2007}).
\newblock
  \urlprefix\url{https://www.sciencedirect.com/science/article/abs/pii/S016501730600083X?via\%3Dihub}.

\bibitem{jensen_shaping_2010}
\bibinfo{author}{Jensen, O.} \& \bibinfo{author}{Mazaheri, A.}
\newblock \bibinfo{title}{Shaping {Functional} {Architecture} by {Oscillatory}
  {Alpha} {Activity}: {Gating} by {Inhibition}}.
\newblock \emph{\bibinfo{journal}{Frontiers in Human Neuroscience}}
  \textbf{\bibinfo{volume}{4}} (\bibinfo{year}{2010}).
\newblock
  \urlprefix\url{https://www.frontiersin.org/articles/10.3389/fnhum.2010.00186/full#B28}.
\newblock \bibinfo{note}{Publisher: Frontiers}.

\bibitem{haegens_-oscillations_2011}
\bibinfo{author}{Haegens, S.}, \bibinfo{author}{Nácher, V.},
  \bibinfo{author}{Luna, R.}, \bibinfo{author}{Romo, R.} \&
  \bibinfo{author}{Jensen, O.}
\newblock \bibinfo{title}{$\alpha$-{Oscillations} in the monkey sensorimotor
  network influence discrimination performance by rhythmical inhibition of
  neuronal spiking}.
\newblock \emph{\bibinfo{journal}{Proceedings of the National Academy of
  Sciences of the United States of America}} \textbf{\bibinfo{volume}{108}},
  \bibinfo{pages}{19377--19382} (\bibinfo{year}{2011}).
\newblock
  \urlprefix\url{https://www.pnas.org/content/early/2011/11/09/1117190108}.

\bibitem{spitzer_beyond_2017}
\bibinfo{author}{Spitzer, B.} \& \bibinfo{author}{Haegens, S.}
\newblock \bibinfo{title}{Beyond the {Status} {Quo}: {A} {Role} for {Beta}
  {Oscillations} in {Endogenous} {Content} ({Re}){Activation}}.
\newblock \emph{\bibinfo{journal}{eNeuro}} \textbf{\bibinfo{volume}{4}}
  (\bibinfo{year}{2017}).
\newblock
  \urlprefix\url{https://www.ncbi.nlm.nih.gov/pmc/articles/PMC5539431/}.

\bibitem{engel_beta-band_2010}
\bibinfo{author}{Engel, A.~K.} \& \bibinfo{author}{Fries, P.}
\newblock \bibinfo{title}{Beta-band oscillations--signalling the status quo?}
\newblock \emph{\bibinfo{journal}{Current Opinion in Neurobiology}}
  \textbf{\bibinfo{volume}{20}}, \bibinfo{pages}{156--165}
  (\bibinfo{year}{2010}).
\newblock
  \urlprefix\url{https://www.sciencedirect.com/science/article/abs/pii/S0959438810000395?via\%3Dihub}.

\bibitem{donner_population_2007}
\bibinfo{author}{Donner, T.~H.} \emph{et~al.}
\newblock \bibinfo{title}{Population activity in the human dorsal pathway
  predicts the accuracy of visual motion detection}.
\newblock \emph{\bibinfo{journal}{Journal of Neurophysiology}}
  \textbf{\bibinfo{volume}{98}}, \bibinfo{pages}{345--359}
  (\bibinfo{year}{2007}).
\newblock
  \urlprefix\url{https://journals.physiology.org/doi/full/10.1152/jn.01141.2006}.

\bibitem{piantoni_beta_2010}
\bibinfo{author}{Piantoni, G.}, \bibinfo{author}{Kline, K.~A.} \&
  \bibinfo{author}{Eagleman, D.~M.}
\newblock \bibinfo{title}{Beta oscillations correlate with the probability of
  perceiving rivalrous visual stimuli}.
\newblock \emph{\bibinfo{journal}{Journal of Vision}}
  \textbf{\bibinfo{volume}{10}}, \bibinfo{pages}{18} (\bibinfo{year}{2010}).
\newblock
  \urlprefix\url{https://jov.arvojournals.org/article.aspx?articleid=2191657}.

\bibitem{siegel_phase-dependent_2009}
\bibinfo{author}{Siegel, M.}, \bibinfo{author}{Warden, M.~R.} \&
  \bibinfo{author}{Miller, E.~K.}
\newblock \bibinfo{title}{Phase-dependent neuronal coding of objects in
  short-term memory}.
\newblock \emph{\bibinfo{journal}{Proceedings of the National Academy of
  Sciences of the United States of America}} \textbf{\bibinfo{volume}{106}},
  \bibinfo{pages}{21341--21346} (\bibinfo{year}{2009}).
\newblock \urlprefix\url{https://www.pnas.org/content/106/50/21341}.

\bibitem{muthuraman_beamformer_2014}
\bibinfo{author}{Muthuraman, M.} \emph{et~al.}
\newblock \bibinfo{title}{Beamformer source analysis and connectivity on
  concurrent {EEG} and {MEG} data during voluntary movements}.
\newblock \emph{\bibinfo{journal}{PloS One}} \textbf{\bibinfo{volume}{9}}
  (\bibinfo{year}{2014}).
\newblock \urlprefix\url{http://dx.doi.org/10.1371/journal.pone.0091441}.

\bibitem{coquelet_comparing_2020}
\bibinfo{author}{Coquelet, N.} \emph{et~al.}
\newblock \bibinfo{title}{Comparing {MEG} and high-density {EEG} for intrinsic
  functional connectivity mapping}.
\newblock \emph{\bibinfo{journal}{Neuroimage}} \bibinfo{pages}{116556}
  (\bibinfo{year}{2020}).
\newblock
  \urlprefix\url{http://www.sciencedirect.com/science/article/pii/S1053811920300434}.

\bibitem{perdikis_subject-oriented_2014}
\bibinfo{author}{Perdikis, S.}, \bibinfo{author}{Leeb, R.} \&
  \bibinfo{author}{Mill{\'a}n, J. d.~R.}
\newblock \bibinfo{title}{Subject-oriented training for motor imagery
  brain-computer interfaces}.
\newblock \emph{\bibinfo{journal}{Conference Proceedings IEEE Engineering in
  Medicine and Biology Society}} \textbf{\bibinfo{volume}{2014}},
  \bibinfo{pages}{1259--1262} (\bibinfo{year}{2014}).
\newblock \urlprefix\url{http://dx.doi.org/10.1109/EMBC.2014.6943826}.

\bibitem{de_vico_fallani_network_2019}
\bibinfo{author}{De~Vico~Fallani, F.} \& \bibinfo{author}{Bassett, D.~S.}
\newblock \bibinfo{title}{Network neuroscience for optimizing brain-computer
  interfaces}.
\newblock \emph{\bibinfo{journal}{Physics of Life Reviews}}
  (\bibinfo{year}{2019}).
\newblock
  \urlprefix\url{http://www.sciencedirect.com/science/article/pii/S1571064519300016}.

\bibitem{benaroch_are_2019}
\bibinfo{author}{Benaroch, C.}, \bibinfo{author}{Jeunet, C.} \&
  \bibinfo{author}{Lotte, F.}
\newblock \bibinfo{title}{Are users' traits informative enough to
  predict/explain their mental-imagery based {BCI} performances?}
\newblock In \emph{\bibinfo{booktitle}{{GBCIC} 2019}}, \bibinfo{pages}{7}
  (\bibinfo{year}{2019}).

\bibitem{ahn_high_2013}
\bibinfo{author}{Ahn, M.}, \bibinfo{author}{Cho, H.}, \bibinfo{author}{Ahn, S.}
  \& \bibinfo{author}{Jun, S.~C.}
\newblock \bibinfo{title}{High theta and low alpha powers may be indicative of
  {BCI}-illiteracy in motor imagery}.
\newblock \emph{\bibinfo{journal}{PLoS ONE}} \textbf{\bibinfo{volume}{8}},
  \bibinfo{pages}{e80886} (\bibinfo{year}{2013}).
\newblock \urlprefix\url{http://dx.doi.org/10.1371/journal.pone.0080886}.

\bibitem{jeunet_predicting_2015}
\bibinfo{author}{Jeunet, C.}, \bibinfo{author}{N'Kaoua, B.},
  \bibinfo{author}{Subramanian, S.}, \bibinfo{author}{Hachet, M.} \&
  \bibinfo{author}{Lotte, F.}
\newblock \bibinfo{title}{Predicting {Mental} {Imagery}-{Based} {BCI}
  {Performance} from {Personality}, {Cognitive} {Profile} and
  {Neurophysiological} {Patterns}}.
\newblock \emph{\bibinfo{journal}{PLoS ONE}} \textbf{\bibinfo{volume}{10}},
  \bibinfo{pages}{e0143962} (\bibinfo{year}{2015}).
\newblock \urlprefix\url{http://dx.doi.org/10.1371/journal.pone.0143962}.

\bibitem{sugata_alpha_2014}
\bibinfo{author}{Sugata, H.} \emph{et~al.}
\newblock \bibinfo{title}{Alpha band functional connectivity correlates with
  the performance of brain-machine interfaces to decode real and imagined
  movements}.
\newblock \emph{\bibinfo{journal}{Frontiers in Human Neuroscience}}
  \textbf{\bibinfo{volume}{8}} (\bibinfo{year}{2014}).
\newblock \urlprefix\url{https://doi.org/10.3389%2Ffnhum.2014.00620}.

\bibitem{de_vico_fallani_multiscale_2013}
\bibinfo{author}{De~Vico~Fallani, F.} \emph{et~al.}
\newblock \bibinfo{title}{Multiscale topological properties of functional brain
  networks during motor imagery after stroke}.
\newblock \emph{\bibinfo{journal}{Neuroimage}} \textbf{\bibinfo{volume}{83}},
  \bibinfo{pages}{438--449} (\bibinfo{year}{2013}).
\newblock \urlprefix\url{http://dx.doi.org/10.1016/j.neuroimage.2013.06.039}.

\bibitem{li_probabilistic_2014}
\bibinfo{author}{Li, J.} \emph{et~al.}
\newblock \bibinfo{title}{Probabilistic diffusion tractography reveals
  improvement of structural network in musicians}.
\newblock \emph{\bibinfo{journal}{PLoS ONE}} \textbf{\bibinfo{volume}{9}},
  \bibinfo{pages}{e105508} (\bibinfo{year}{2014}).
\newblock
  \urlprefix\url{https://journals.plos.org/plosone/article?id=10.1371/journal.pone.0105508}.

\bibitem{calmels_neural_2020}
\bibinfo{author}{Calmels, C.}
\newblock \bibinfo{title}{Neural correlates of motor expertise: {Extensive}
  motor training and cortical changes}.
\newblock \emph{\bibinfo{journal}{Brain Research}}
  \textbf{\bibinfo{volume}{1739}}, \bibinfo{pages}{146323}
  (\bibinfo{year}{2020}).
\newblock
  \urlprefix\url{http://www.sciencedirect.com/science/article/pii/S0006899319303695}.

\bibitem{hetu_neural_2013}
\bibinfo{author}{H{\'e}tu, S.} \emph{et~al.}
\newblock \bibinfo{title}{The neural network of motor imagery: an {ALE}
  meta-analysis}.
\newblock \emph{\bibinfo{journal}{Neuroscience \& Biobehavioral Reviews}}
  \textbf{\bibinfo{volume}{37}}, \bibinfo{pages}{930--949}
  (\bibinfo{year}{2013}).
\newblock \urlprefix\url{http://dx.doi.org/10.1016/j.neubiorev.2013.03.017}.

\bibitem{hardwick_neural_2018}
\bibinfo{author}{Hardwick, R.~M.}, \bibinfo{author}{Caspers, S.},
  \bibinfo{author}{Eickhoff, S.~B.} \& \bibinfo{author}{Swinnen, S.~P.}
\newblock \bibinfo{title}{Neural correlates of action: {Comparing}
  meta-analyses of imagery, observation, and execution}.
\newblock \emph{\bibinfo{journal}{Neuroscience \& Biobehavioral Reviews}}
  \textbf{\bibinfo{volume}{94}}, \bibinfo{pages}{31--44}
  (\bibinfo{year}{2018}).
\newblock
  \urlprefix\url{http://www.sciencedirect.com/science/article/pii/S0149763417309284}.

\bibitem{dayan_neuroplasticity_2011}
\bibinfo{author}{Dayan, E.} \& \bibinfo{author}{Cohen, L.~G.}
\newblock \bibinfo{title}{Neuroplasticity subserving motor skill learning}.
\newblock \emph{\bibinfo{journal}{Neuron}} \textbf{\bibinfo{volume}{72}},
  \bibinfo{pages}{443--454} (\bibinfo{year}{2011}).
\newblock
  \urlprefix\url{https://www.ncbi.nlm.nih.gov/pmc/articles/PMC3217208/}.

\bibitem{van_zomeren_clinical_1994}
\bibinfo{author}{van Zomeren, A.~H.} \& \bibinfo{author}{Brouwer, W.~H.}
\newblock \emph{\bibinfo{title}{Clinical neuropsychology of attention}}.
\newblock Clinical neuropsychology of attention (\bibinfo{publisher}{Oxford
  University Press}, \bibinfo{address}{New York, NY, US},
  \bibinfo{year}{1994}).

\bibitem{wolpert_motor_2012}
\bibinfo{author}{Wolpert, D.~M.} \& \bibinfo{author}{Landy, M.~S.}
\newblock \bibinfo{title}{Motor {Control} is {Decision}-{Making}}.
\newblock \emph{\bibinfo{journal}{Current Opinion in Neurobiology}}
  \textbf{\bibinfo{volume}{22}}, \bibinfo{pages}{996--1003}
  (\bibinfo{year}{2012}).
\newblock
  \urlprefix\url{https://www.ncbi.nlm.nih.gov/pmc/articles/PMC3434279/}.

\bibitem{guillot_neurophysiological_2010}
\bibinfo{editor}{Guillot, A.} \& \bibinfo{editor}{Collet, C.} (eds.)
  \emph{\bibinfo{title}{The neurophysiological foundations of mental and motor
  imagery}} (\bibinfo{publisher}{Oxford University Press},
  \bibinfo{address}{Oxford, New York}, \bibinfo{year}{2010}).

\bibitem{wulf_attention_2007}
\bibinfo{author}{Wulf, G.}
\newblock \emph{\bibinfo{title}{Attention and motor skill learning}}.
\newblock Attention and motor skill learning (\bibinfo{publisher}{Human
  Kinetics}, \bibinfo{address}{Champaign, IL, US}, \bibinfo{year}{2007}).

\bibitem{lohse_role_2014}
\bibinfo{author}{Lohse, K.~R.}, \bibinfo{author}{Jones, M.},
  \bibinfo{author}{Healy, A.~F.} \& \bibinfo{author}{Sherwood, D.~E.}
\newblock \bibinfo{title}{The role of attention in motor control}.
\newblock \emph{\bibinfo{journal}{Journal of Experimental Psychology: General}}
  \textbf{\bibinfo{volume}{143}}, \bibinfo{pages}{930--948}
  (\bibinfo{year}{2014}).
\newblock \urlprefix\url{http://dx.doi.org/10.1037/a0032817}.

\bibitem{dayan_learning_2000}
\bibinfo{author}{Dayan, P.}, \bibinfo{author}{Kakade, S.} \&
  \bibinfo{author}{Montague, P.~R.}
\newblock \bibinfo{title}{Learning and selective attention}.
\newblock \emph{\bibinfo{journal}{Nature Neuroscience}}
  \textbf{\bibinfo{volume}{3}}, \bibinfo{pages}{1218--1223}
  (\bibinfo{year}{2000}).
\newblock \urlprefix\url{https://www.nature.com/articles/nn1100_1218}.

\bibitem{gottlieb_attention_2012}
\bibinfo{author}{Gottlieb, J.}
\newblock \bibinfo{title}{Attention, {Learning}, and the {Value} of
  {Information}}.
\newblock \emph{\bibinfo{journal}{Neuron}} \textbf{\bibinfo{volume}{76}},
  \bibinfo{pages}{281--295} (\bibinfo{year}{2012}).
\newblock
  \urlprefix\url{http://www.sciencedirect.com/science/article/pii/S0896627312008884}.

\bibitem{yin_dynamic_2009}
\bibinfo{author}{Yin, H.~H.} \emph{et~al.}
\newblock \bibinfo{title}{Dynamic reorganization of striatal circuits during
  the acquisition and consolidation of a skill}.
\newblock \emph{\bibinfo{journal}{Nature Neuroscience}}
  \textbf{\bibinfo{volume}{12}}, \bibinfo{pages}{333--341}
  (\bibinfo{year}{2009}).
\newblock
  \urlprefix\url{https://www.ncbi.nlm.nih.gov/pmc/articles/PMC2774785/}.

\bibitem{boto_measurements_2017}
\bibinfo{author}{Boto, E.} \emph{et~al.}
\newblock \bibinfo{title}{A new generation of magnetoencephalography: Room
  temperature measurements using optically-pumped magnetometers}.
\newblock \emph{\bibinfo{journal}{NeuroImage}} \textbf{\bibinfo{volume}{149}},
  \bibinfo{pages}{404 -- 414} (\bibinfo{year}{2017}).
\newblock
  \urlprefix\url{http://www.sciencedirect.com/science/article/pii/S1053811917300411}.

\bibitem{labyt_magnetoencephalography_2018}
\bibinfo{author}{Labyt, E.} \emph{et~al.}
\newblock \bibinfo{title}{Magnetoencephalography with optically pumped {4He}
  magnetometers at ambient temperature}.
\newblock \emph{\bibinfo{journal}{IEEE Transactions on Medical Imaging}}
  (\bibinfo{year}{2018}).
\newblock \urlprefix\url{https://ieeexplore.ieee.org/document/8411463}.

\bibitem{boto_wearable_2019}
\bibinfo{author}{Boto, E.} \emph{et~al.}
\newblock \bibinfo{title}{Wearable neuroimaging: {Combining} and contrasting
  magnetoencephalography and electroencephalography}.
\newblock \emph{\bibinfo{journal}{Neuroimage}} \bibinfo{pages}{116099}
  (\bibinfo{year}{2019}).
\newblock
  \urlprefix\url{http://www.sciencedirect.com/science/article/pii/S1053811919306901}.

\bibitem{tierney_optically_2019}
\bibinfo{author}{Tierney, T.~M.} \emph{et~al.}
\newblock \bibinfo{title}{Optically pumped magnetometers: {From} quantum
  origins to multi-channel magnetoencephalography}.
\newblock \emph{\bibinfo{journal}{Neuroimage}}  (\bibinfo{year}{2019}).
\newblock
  \urlprefix\url{http://www.sciencedirect.com/science/article/pii/S1053811919304550}.

\bibitem{mitchell2013gendered}
\bibinfo{author}{Mitchell, S.~M.}, \bibinfo{author}{Lange, S.} \&
  \bibinfo{author}{Brus, H.}
\newblock \bibinfo{title}{Gendered citation patterns in international relations
  journals}.
\newblock \emph{\bibinfo{journal}{International Studies Perspectives}}
  \textbf{\bibinfo{volume}{14}}, \bibinfo{pages}{485--492}
  (\bibinfo{year}{2013}).

\bibitem{dion2018gendered}
\bibinfo{author}{Dion, M.~L.}, \bibinfo{author}{Sumner, J.~L.} \&
  \bibinfo{author}{Mitchell, S.~M.}
\newblock \bibinfo{title}{Gendered citation patterns across political science
  and social science methodology fields}.
\newblock \emph{\bibinfo{journal}{Political Analysis}}
  \textbf{\bibinfo{volume}{26}}, \bibinfo{pages}{312--327}
  (\bibinfo{year}{2018}).

\bibitem{caplar2017quantitative}
\bibinfo{author}{Caplar, N.}, \bibinfo{author}{Tacchella, S.} \&
  \bibinfo{author}{Birrer, S.}
\newblock \bibinfo{title}{Quantitative evaluation of gender bias in
  astronomical publications from citation counts}.
\newblock \emph{\bibinfo{journal}{Nature Astronomy}}
  \textbf{\bibinfo{volume}{1}}, \bibinfo{pages}{0141} (\bibinfo{year}{2017}).

\bibitem{maliniak2013gender}
\bibinfo{author}{Maliniak, D.}, \bibinfo{author}{Powers, R.} \&
  \bibinfo{author}{Walter, B.~F.}
\newblock \bibinfo{title}{The gender citation gap in international relations}.
\newblock \emph{\bibinfo{journal}{International Organization}}
  \textbf{\bibinfo{volume}{67}}, \bibinfo{pages}{889--922}
  (\bibinfo{year}{2013}).

\bibitem{Dworkin2020.01.03.894378}
\bibinfo{author}{Dworkin, J.~D.} \emph{et~al.}
\newblock \bibinfo{title}{The extent and drivers of gender imbalance in
  neuroscience reference lists}.
\newblock \emph{\bibinfo{journal}{bioRxiv}}  (\bibinfo{year}{2020}).
\newblock
  \urlprefix\url{https://www.biorxiv.org/content/early/2020/01/11/2020.01.03.894378}.
\newblock
  \eprint{https://www.biorxiv.org/content/early/2020/01/11/2020.01.03.894378.full.pdf}.

\bibitem{zhou_dale_2020_3672110}
\bibinfo{author}{Zhou, D.} \emph{et~al.}
\newblock \bibinfo{title}{Gender diversity statement and code notebook v1.0}
  (\bibinfo{year}{2020}).
\newblock \urlprefix\url{https://doi.org/10.5281/zenodo.3672110}.

\bibitem{ambekar2009name}
\bibinfo{author}{Ambekar, A.}, \bibinfo{author}{Ward, C.},
  \bibinfo{author}{Mohammed, J.}, \bibinfo{author}{Male, S.} \&
  \bibinfo{author}{Skiena, S.}
\newblock \bibinfo{title}{Name-ethnicity classification from open sources}.
\newblock In \emph{\bibinfo{booktitle}{Proceedings of the 15th ACM SIGKDD
  international conference on Knowledge Discovery and Data Mining}},
  \bibinfo{pages}{49--58} (\bibinfo{year}{2009}).

\bibitem{sood2018predicting}
\bibinfo{author}{Sood, G.} \& \bibinfo{author}{Laohaprapanon, S.}
\newblock \bibinfo{title}{Predicting race and ethnicity from the sequence of
  characters in a name}.
\newblock \emph{\bibinfo{journal}{arXiv preprint arXiv:1805.02109}}
  (\bibinfo{year}{2018}).

\end{thebibliography}

\section*{Acknowledgments}
This work was partially supported by French program ''Investissements d'avenir'' ANR-10-IAIHU-06; ''ANR-NIH CRCNS'' ANR-15-NEUC-0006-02 and by NICHD 1R01HD086888-01. The funders had no role in study design, data collection and analysis, decision to publish, or preparation of the manuscript.
This work was performed on a platform of France Life Imaging network partly funded by the grant "ANR-11-INBS-0006".

\section*{Citation Diversity Statement}
Recent work in several fields of science has identified a bias in citation practices such that papers from women and other minority scholars are under-cited relative to the number of such papers in the field \cite{mitchell2013gendered,dion2018gendered,caplar2017quantitative, maliniak2013gender, Dworkin2020.01.03.894378}. Here we sought to proactively consider choosing references that reflect the diversity of the field in thought, form of contribution, gender, race, ethnicity, and other factors. We used automatic classification of gender based on the first names of the first and last authors \cite{Dworkin2020.01.03.894378,zhou_dale_2020_3672110}, with possible combinations including man/ man, man/woman, woman/man, and woman/woman. Code for this classification is open source and available online \cite{zhou_dale_2020_3672110}. By this measure (and excluding self-citations to the first and last authors of our current paper), our references contain 4.96\% woman(first)/woman(last), 13.24\% man/woman, 16.34\% woman/man, and 65.47\% man/man. Second, we obtained predicted racial/ethnic category of the first and last author of each reference by databases that store the probability of a first and last name being carried by an author of color \cite{ambekar2009name, sood2018predicting}. By this measure (and excluding self-citations), our references contain 12.99\% author of color (first)/author of color(last), 14.83\% white author/author of color, 21.91\% author of color/white author, and 50.26\% white author/white author. We look forward to future work that could help us to better understand how to support equitable practices in science.

\section*{Authors contributions}
MC, DS, NG, LH, SD, DSB and FDVF initiated research; MCC, MC, DS, LH, DSB and FDVF designed research; MCC, DS and LH performed research; MCC, DS, LH and AEK contributed analytic tools; MCC and AEK analyzed data; and MCC, DSB, and FDVF wrote the paper. All authors revised and approved the manuscript.

\section*{Additional Information}
Supplementary Information accompanies this paper.

%\newpage

\selectlanguage{english}
\clearpage
\end{document}

\captionsetup[figure]{name=Figure}
\captionsetup[table]{name=Table}
\renewcommand\thefigure{S\arabic{figure}}
\renewcommand\thetable{S\arabic{table}}

%\newif\iflatexml\latexmlfalse
%\providecommand{\tightlist}{\setlength{\itemsep}{0pt}\setlength{\parskip}{0pt}}%

\title{BCI learning induces core-periphery reorganization in M/EEG multiplex brain networks}
\author{M.-C. Corsi* et al}

\maketitle

\section*{Supplementary Figures}

\begin{figure}[ht!]
\begin{center}
\includegraphics[width=\textwidth]{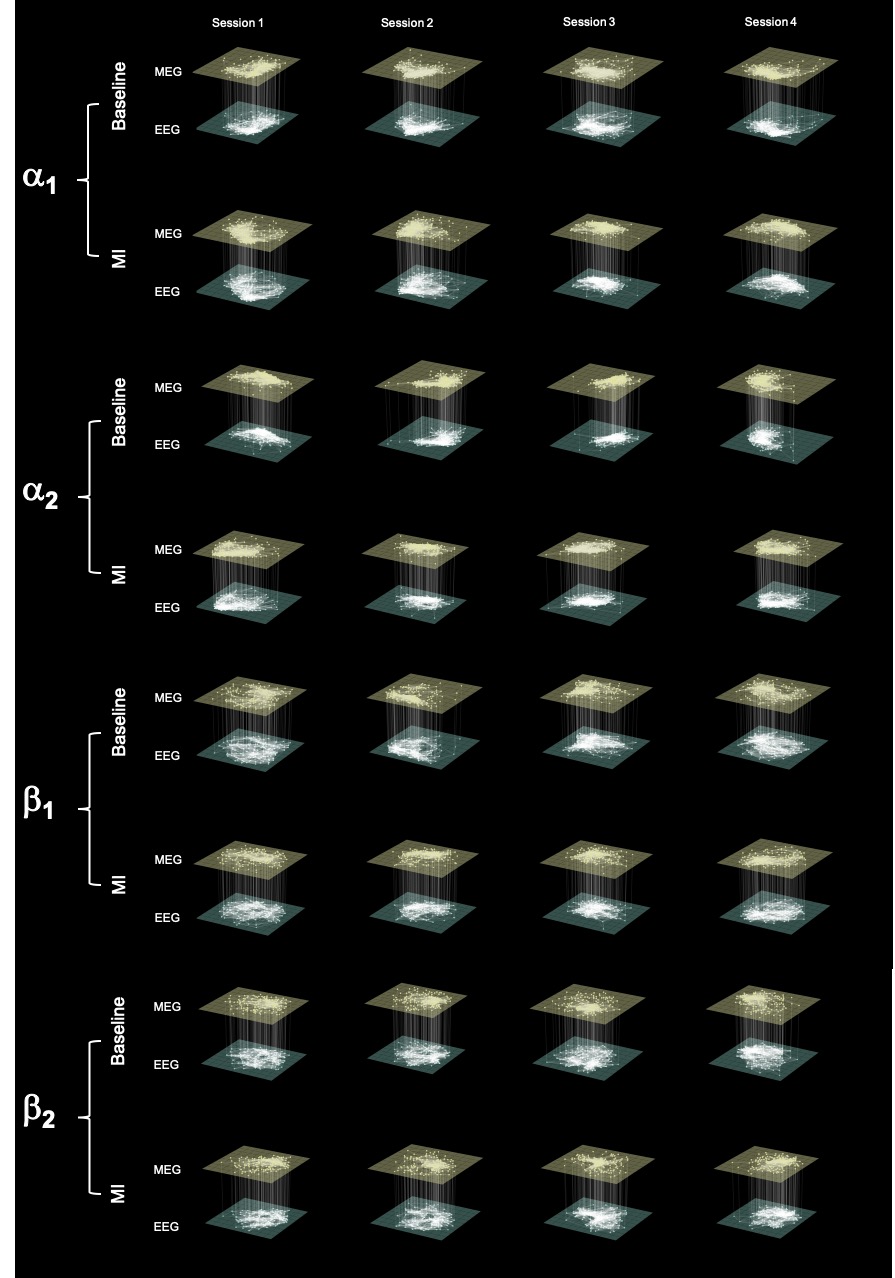}
\caption{Evolution of the single layers for each condition within the $\alpha$ and the $\beta$ ranges (average over the participants).}
\label{Figure1_true}
\end{center}
\end{figure}

\begin{figure}[ht!]
\begin{center}
\includegraphics[width=\textwidth]{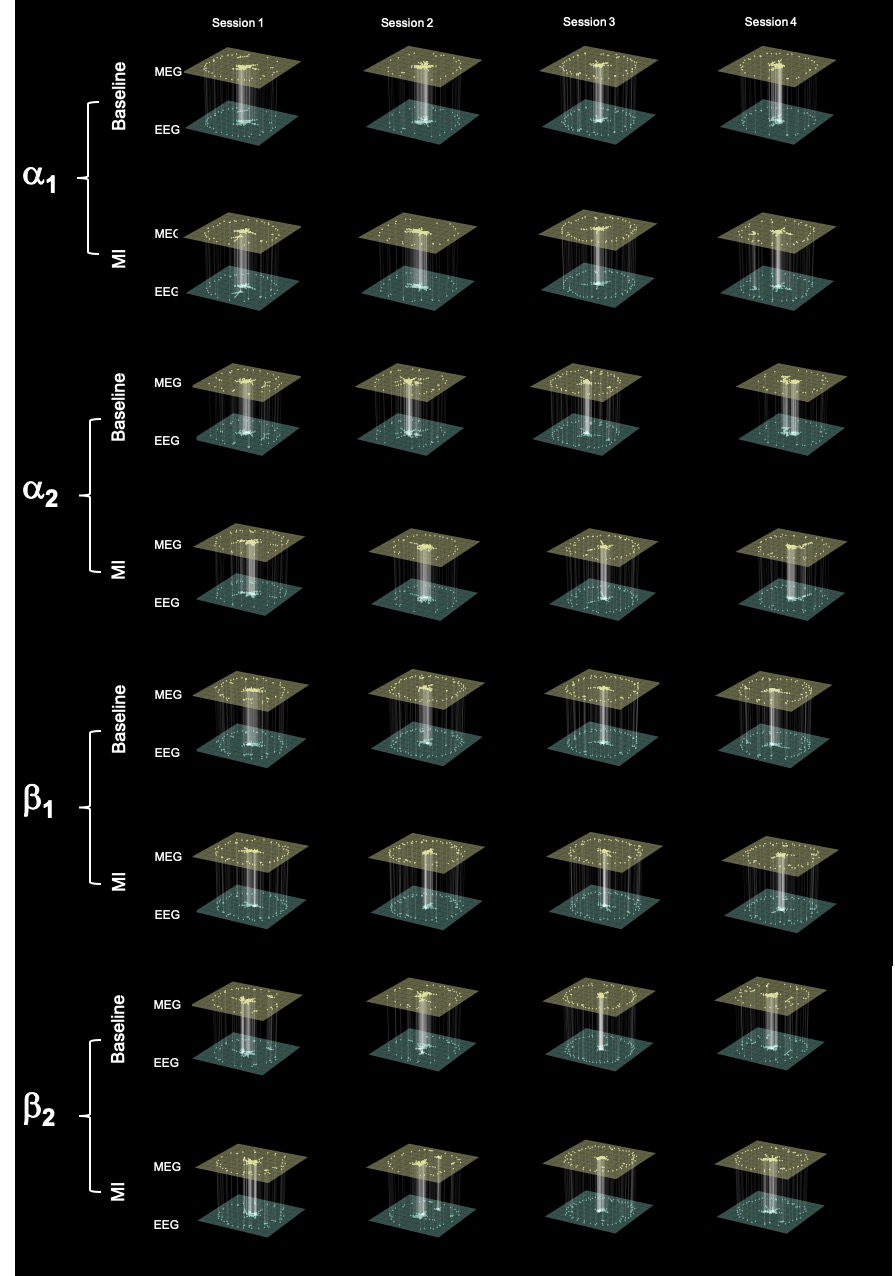}
\caption{Evolution of the single layers for each condition within the $\alpha$ and the $\beta$ ranges (Subject 03).}
\label{Figure2_true}
\end{center}
\end{figure}

\begin{figure}[ht!]
\begin{center}
\includegraphics[width=\textwidth]{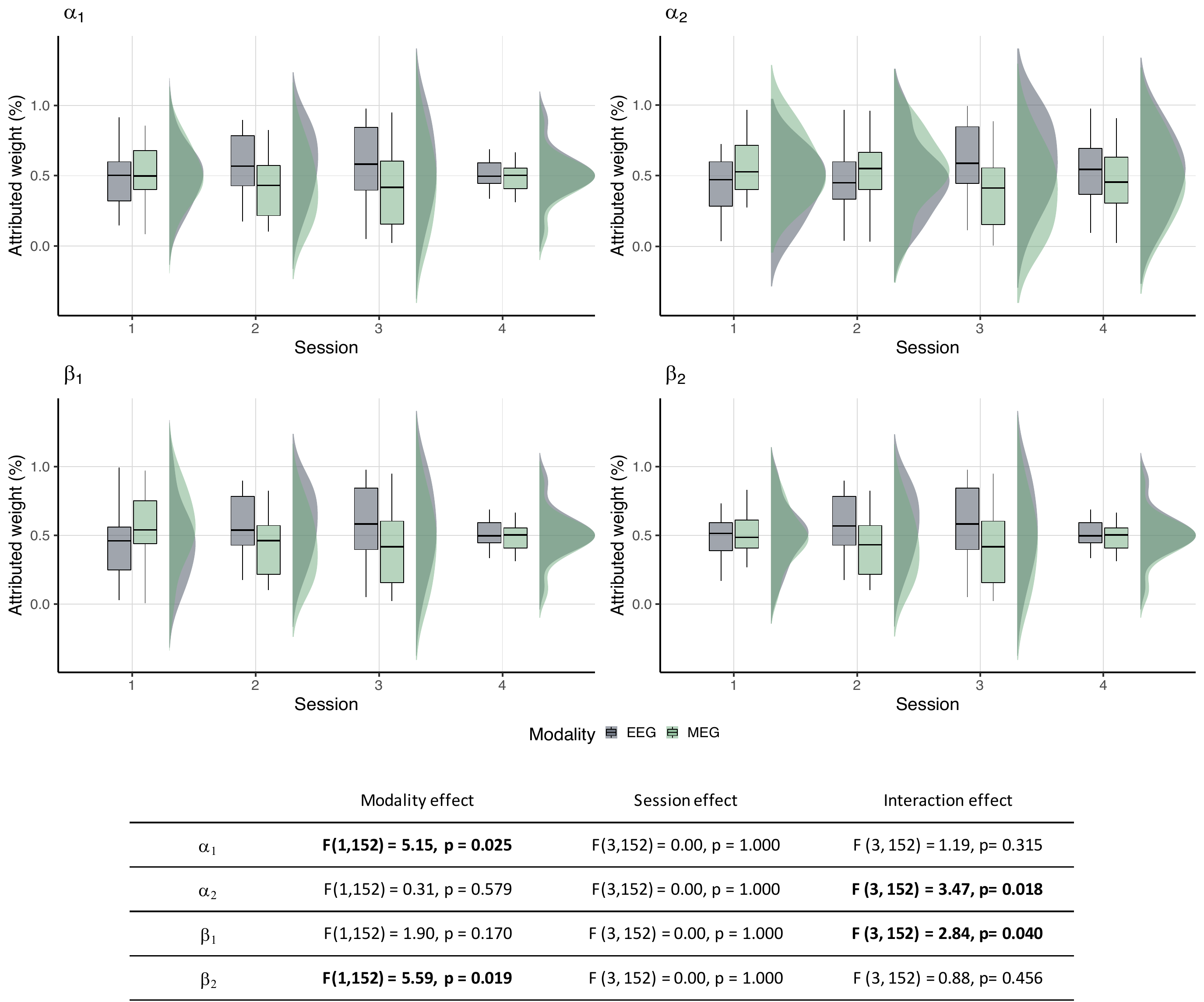}
\caption{Evolution of attributed weights over sessions within the $\alpha$ and the $\beta$ ranges. We performed a two-way ANOVA using permutation tests.}
\label{Figure3}
\end{center}
\end{figure}

\newpage

\begin{figure}
\centering
\includegraphics[width=\textwidth]{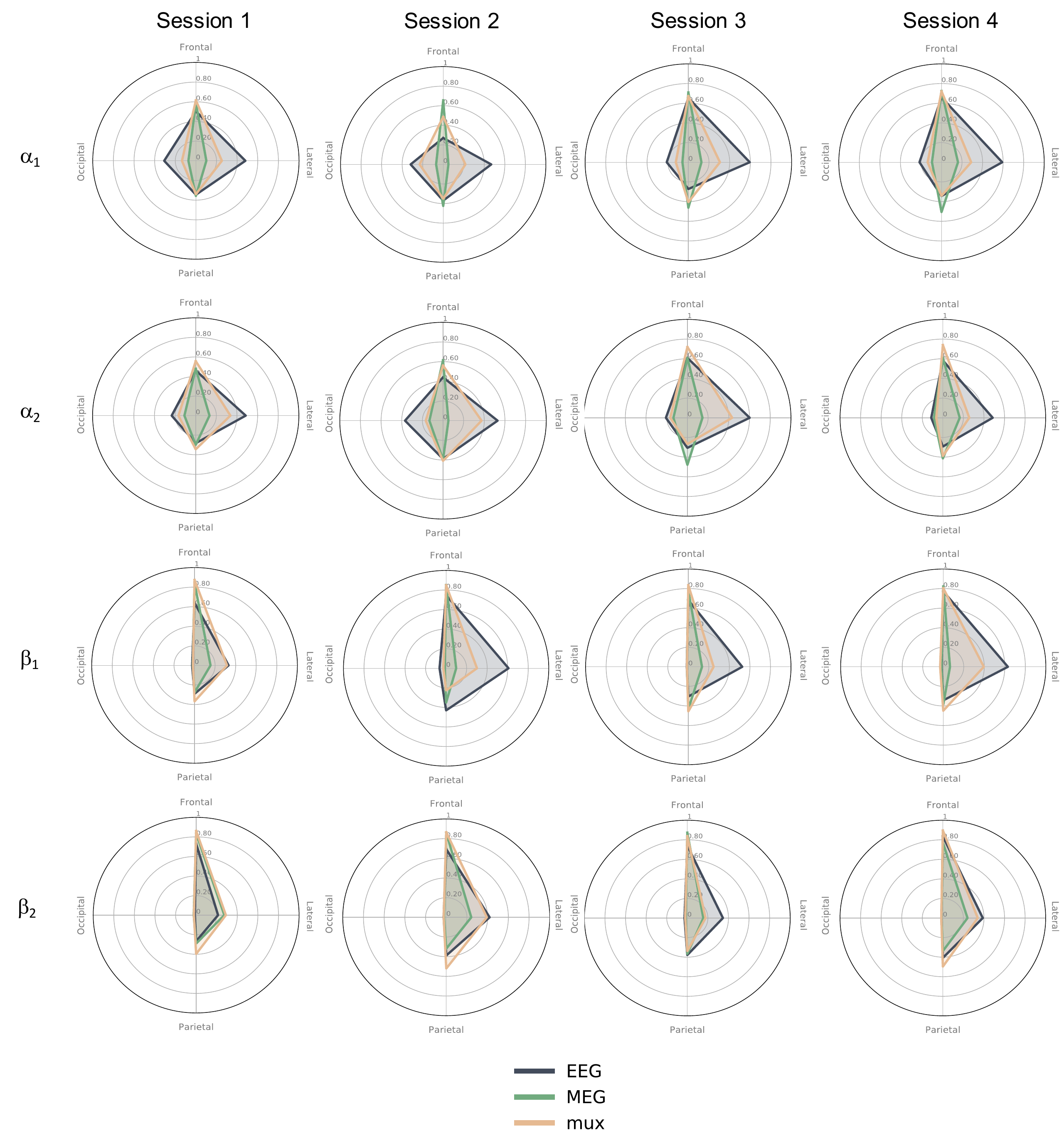}
\caption{Evolution of single layer and multiplex coreness values over the BCI training in MI condition. For a given axis associated with a single brain lobe, we plotted the median coreness value obtained across the subjects and the ROIs that belong to the lobe, respectively in EEG, MEG and multiplex (mux). Each line corresponds to a specific frequency band (respectively $\alpha_1$, $\alpha_2$, $\beta_1$ and $\beta_2$) and each column to a specific session.}
\label{Figure4}
\end{figure}
\newpage

\begin{figure}
\centering
\includegraphics[width=\textwidth]{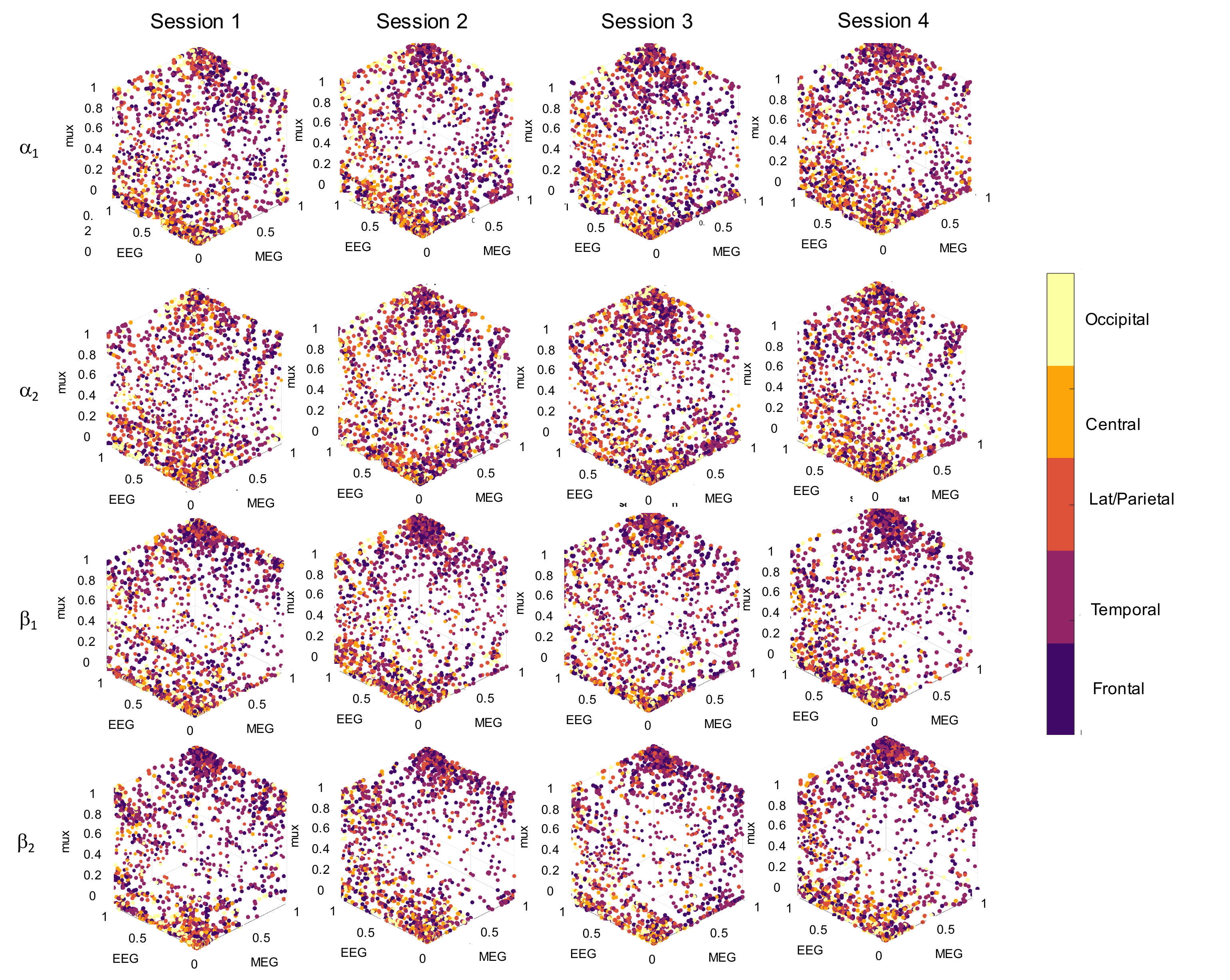}
\caption{Evolution of single layer and multiplex coreness over the sessions in MI condition. For a given point associated with a single ROI and a single subject, its (x,y,z) coordinates correspond to the associated coreness value respectively in EEG, MEG, and multiplex (mux) and its color to its associated brain region. Each line corresponds to a specific frequency band and each column to a given session. We observed a cluster of ROIs (Cl1) that presents a maximal level of coreness in every modality in most of the sessions, both with $\alpha$ and $\beta$ frequency ranges. They mostly belonged to frontal, lateral, and temporal lobes. More specifically, in the $\alpha_{2}$ frequency band, we observed a homogeneous distribution of points at the beginning of training. Then, two main clusters appeared, Cl1 and another, mostly in the plane $C_{MEG}=0$ (Cl2) involving ROIs located in central, latero-parietal, and temporal lobes. In the $\beta_{1}$ frequency band, we obtained a distribution less homogeneous than in $\alpha_{2}$ with Cl1 and the appearance of two other clusters in the plane $C_{MEG}=0$. One is associated with $C_{Mux}$ larger than 0.3 (Cl3) gathering ROIs that belong to central, frontal, and temporal areas. The other (Cl4) is characterized by a minimal level of coreness in every modality associated with ROIs that mainly belong to occipital and central areas.}
\label{Figure5}
\end{figure}
\newpage

\begin{figure}
\centering
\includegraphics[width=\textwidth]{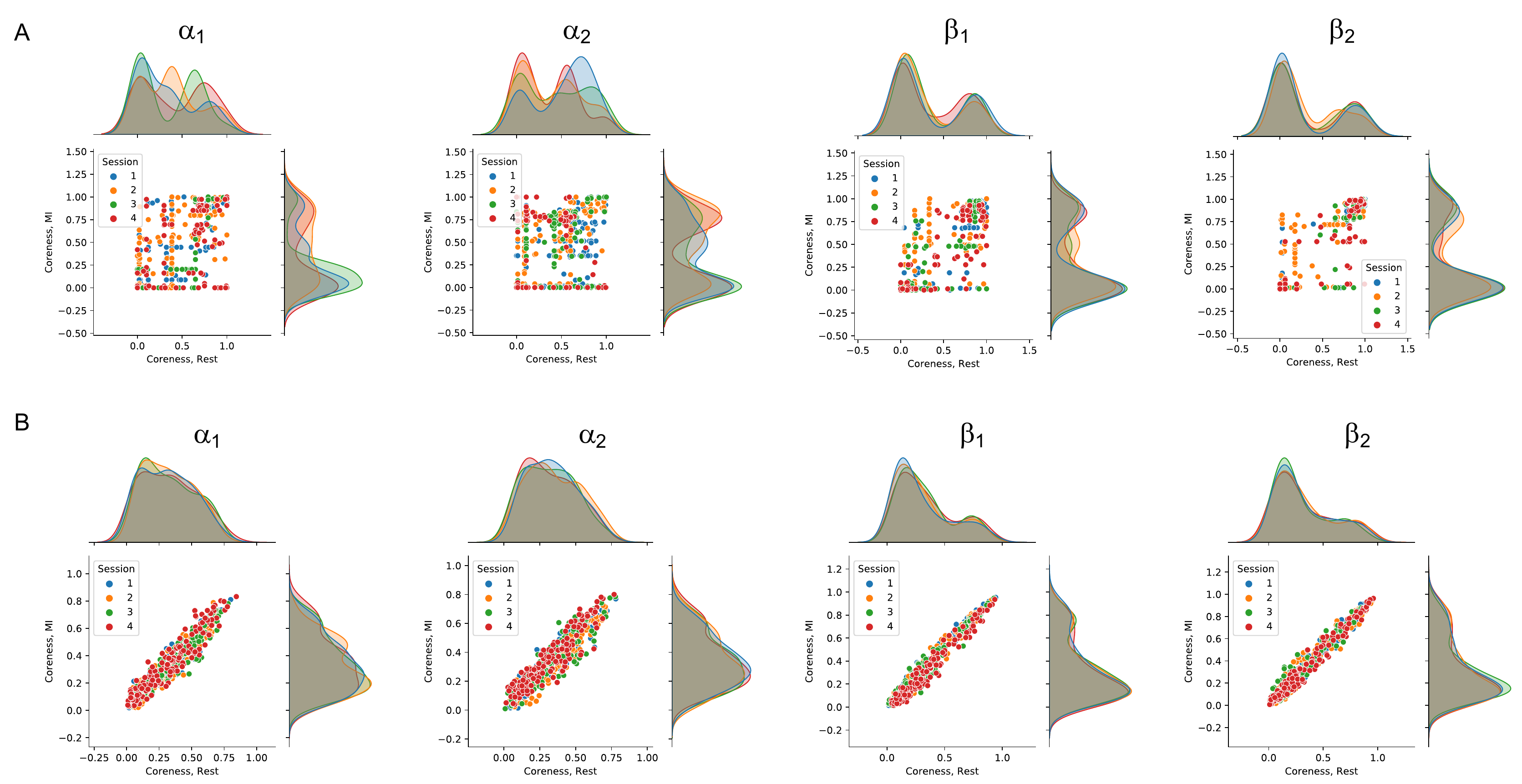}
\caption{Multiplex coreness evolution over the training and the conditions, within the $\alpha$ and the $\beta$ ranges. Each dot corresponds to a given ROI On the X axis are represented the coreness values obtained during the Rest condition; on the Y axis are presented the coreness values obtained during the MI condition. (A) Results obtained from Subject 03. (B) Results obtained from the average performed over the subjects. The strong inter-subject variability engenders a lack of separability between conditions when one considers the average over the subjects.}
\label{Figure6}
\end{figure}
\newpage

\begin{figure}
\centering
\includegraphics[width=\textwidth]{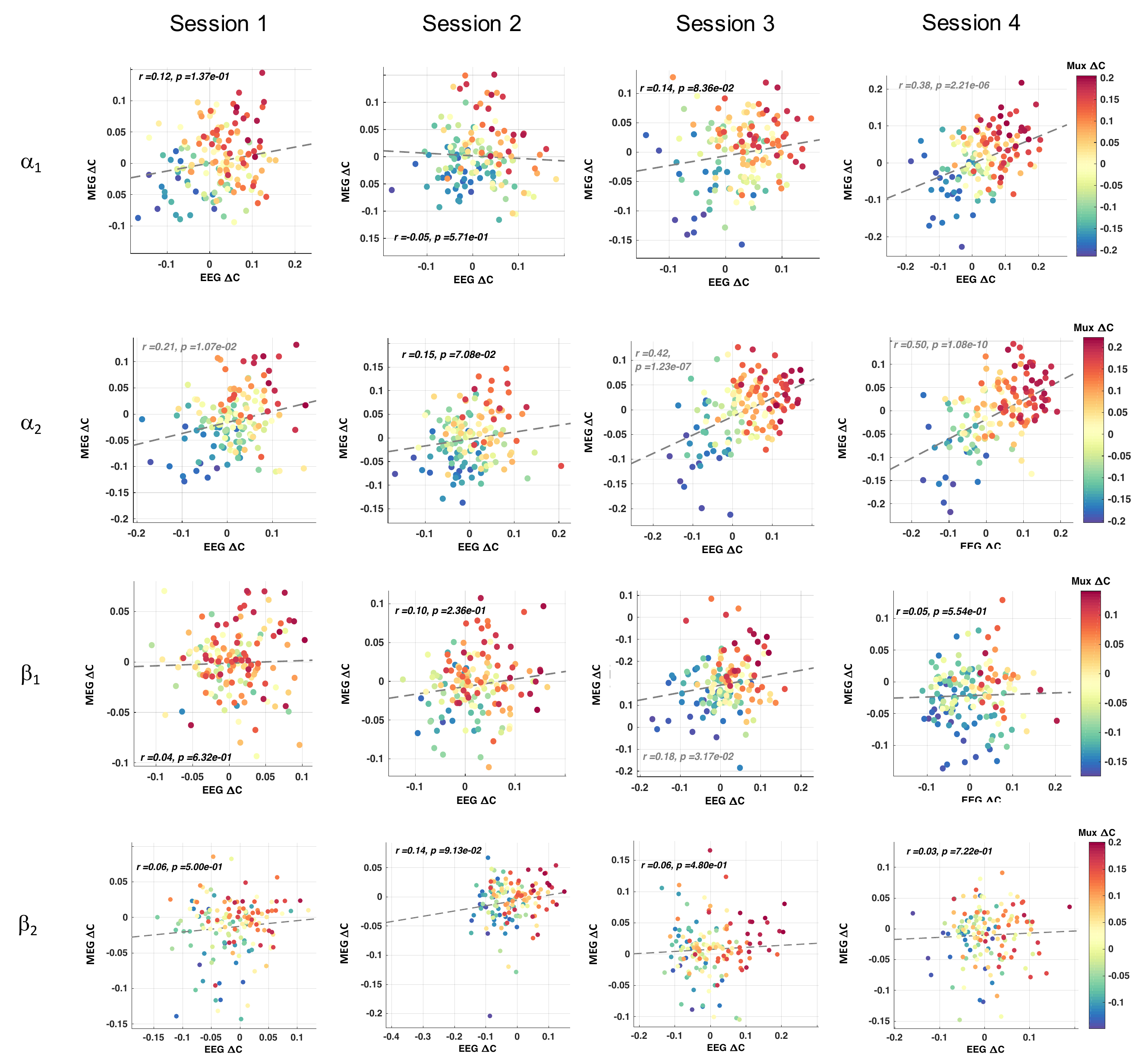}
\caption{Evolution of the averaged relative coreness ($\Delta$C) across the subjects over the training. On the X axis are represented the averaged $\Delta$C values, obtained with the EEG layer, on the Y axis those obtained with the MEG layer and the color of the markers is associated with the values obtained with mux. Each line corresponds to a specific frequency band and each column to a given session.}
\label{Figure7}
\end{figure}
\newpage

\begin{figure}
\centering
\includegraphics[scale=0.51]{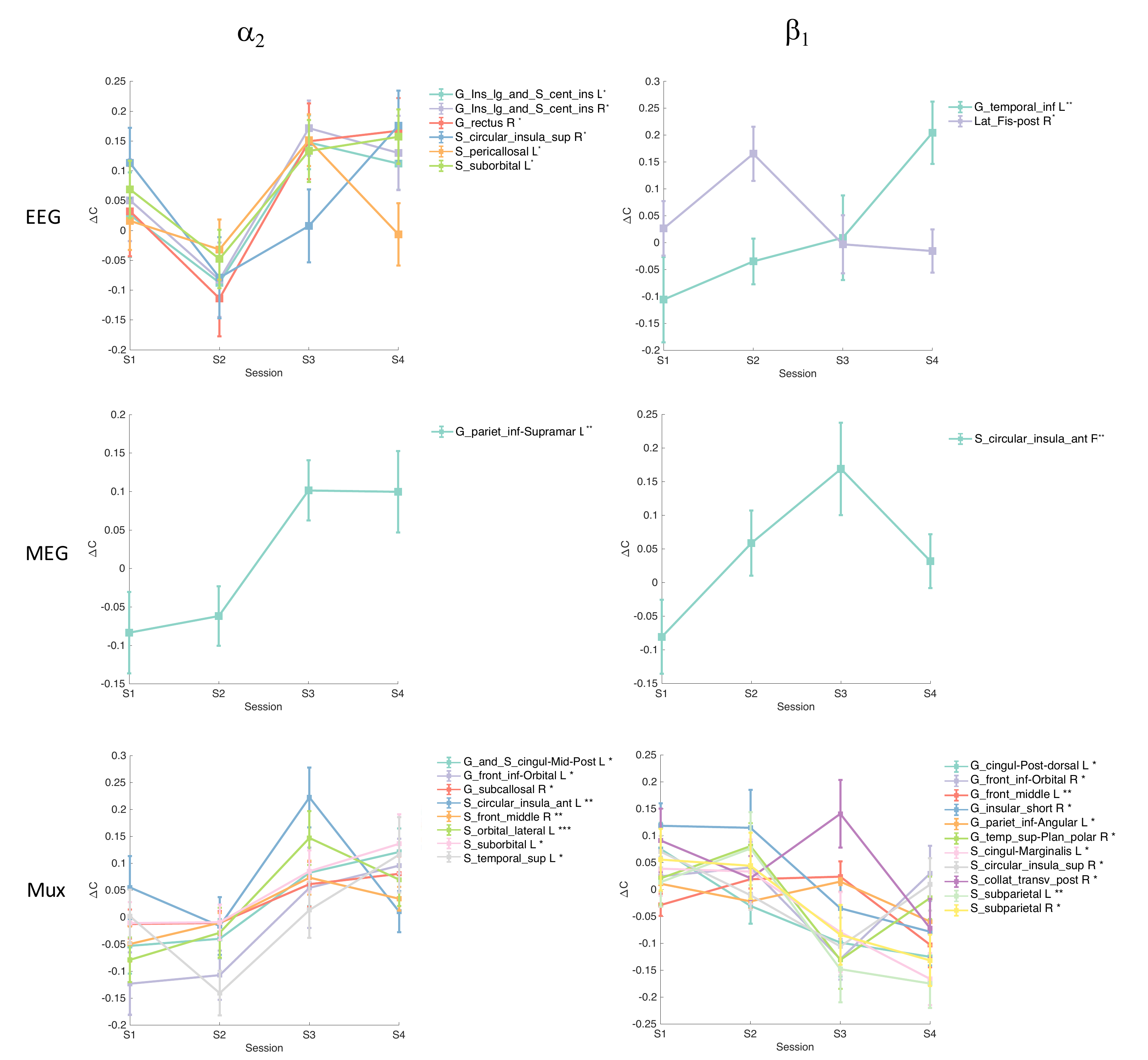}
\caption{Distribution of relative coreness over the training in the $\alpha_2$ and $\beta_1$ frequency bands. Only the ROIs that present a significant session effect are represented (one-way ANOVA, $p < 0.05$). Modalities were considered separately.}
\label{Figure8}
\end{figure}
\newpage

\begin{figure}
\centering
\includegraphics[width=\textwidth]{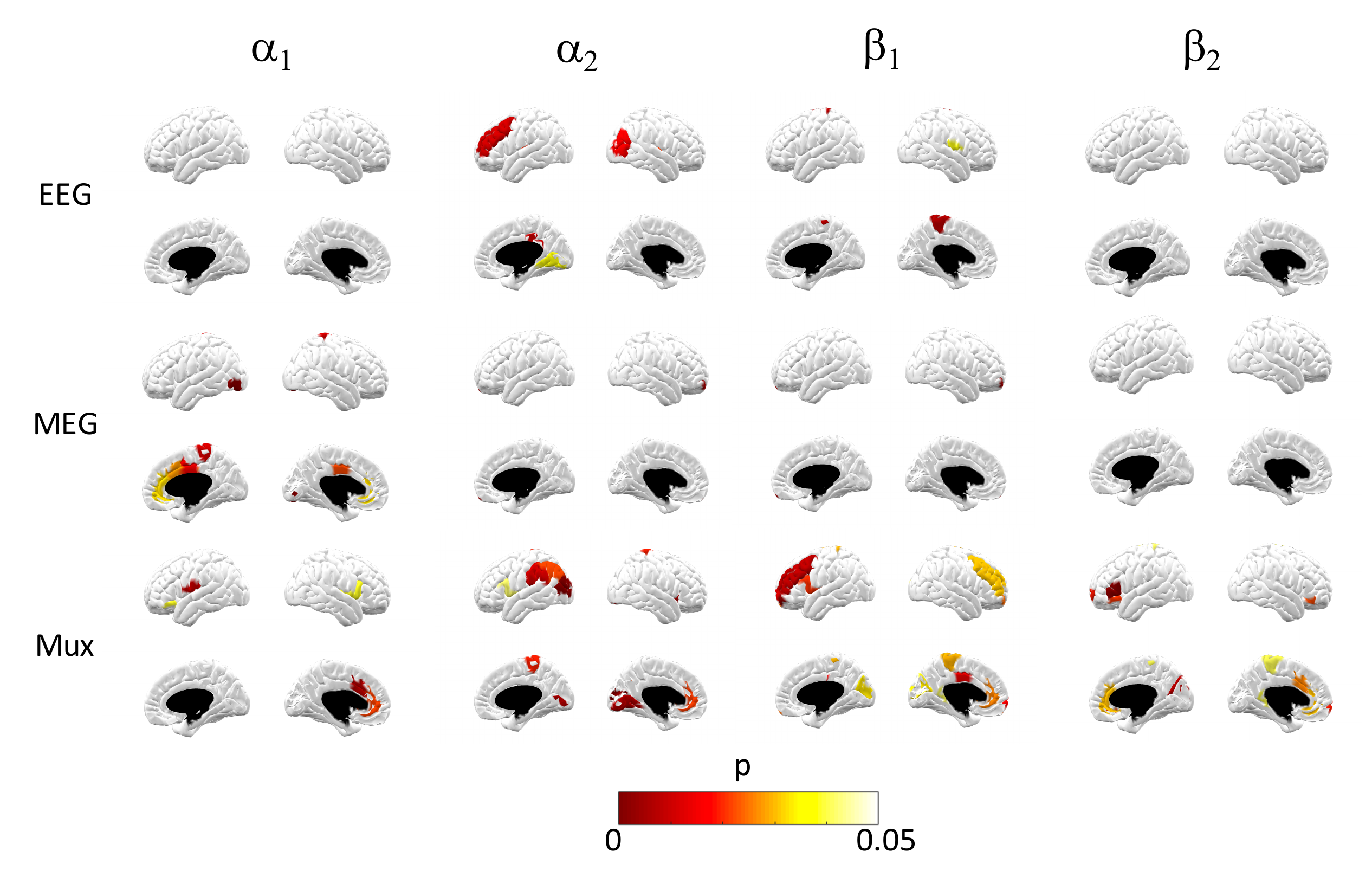}
\caption{Significant session effect obtained with relative coreness (one-way ANOVA, $p < 0.05$) over sessions in the $\alpha$ and $\beta$ frequency ranges. Modalities were considered separately.}
\label{Figure9}
\end{figure}
\newpage

\begin{figure}
\centering
\includegraphics[width=\textwidth]{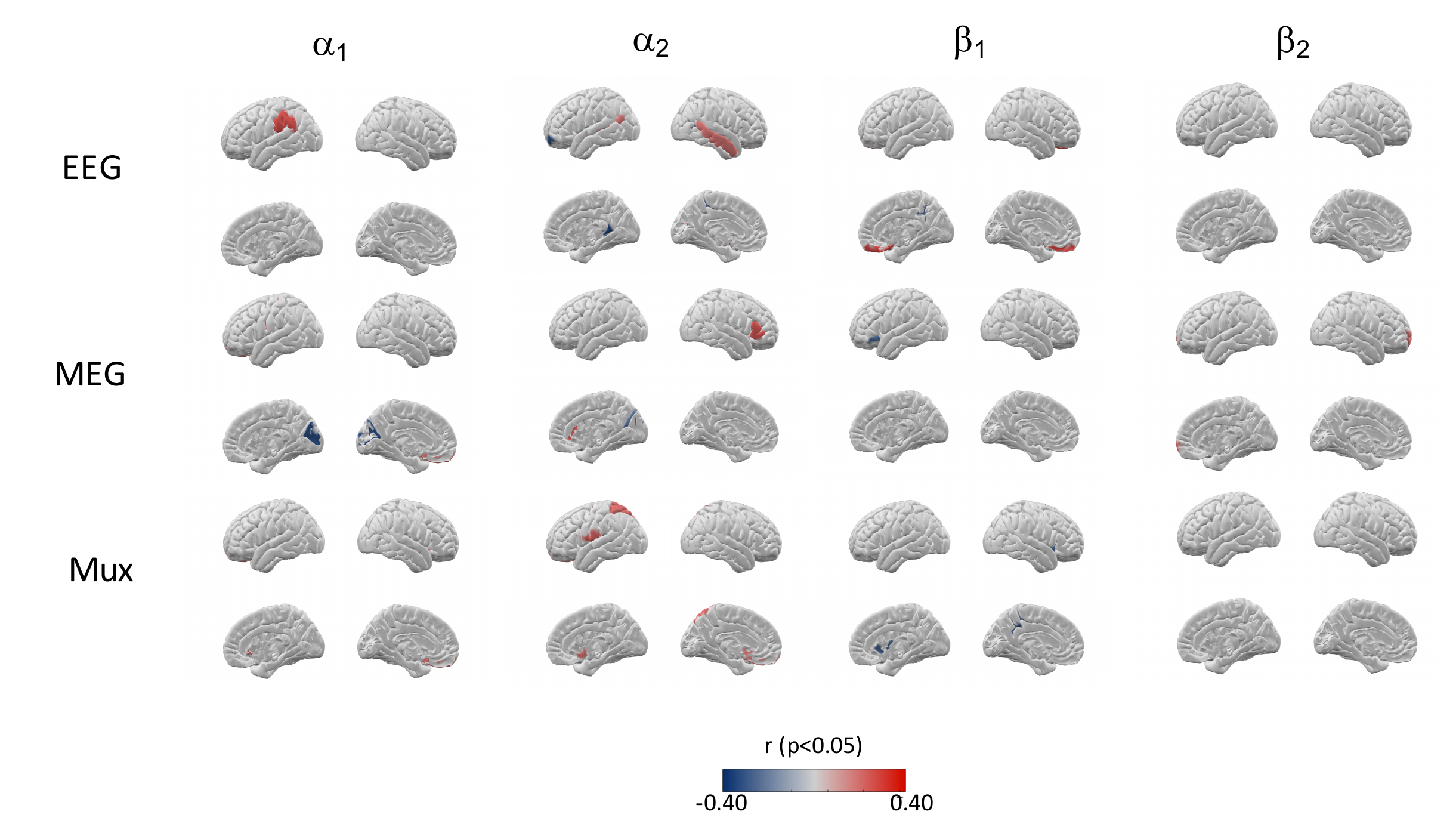} 
\caption{Repeated correlations between relative coreness and BCI scores in EEG, MEG, and mux ($p < 0.05$) within the $\alpha$ and the $\beta$ frequency ranges.}
\label{Figure10}
\end{figure}
\newpage

\begin{figure}
\centering
\includegraphics[width=\textwidth]{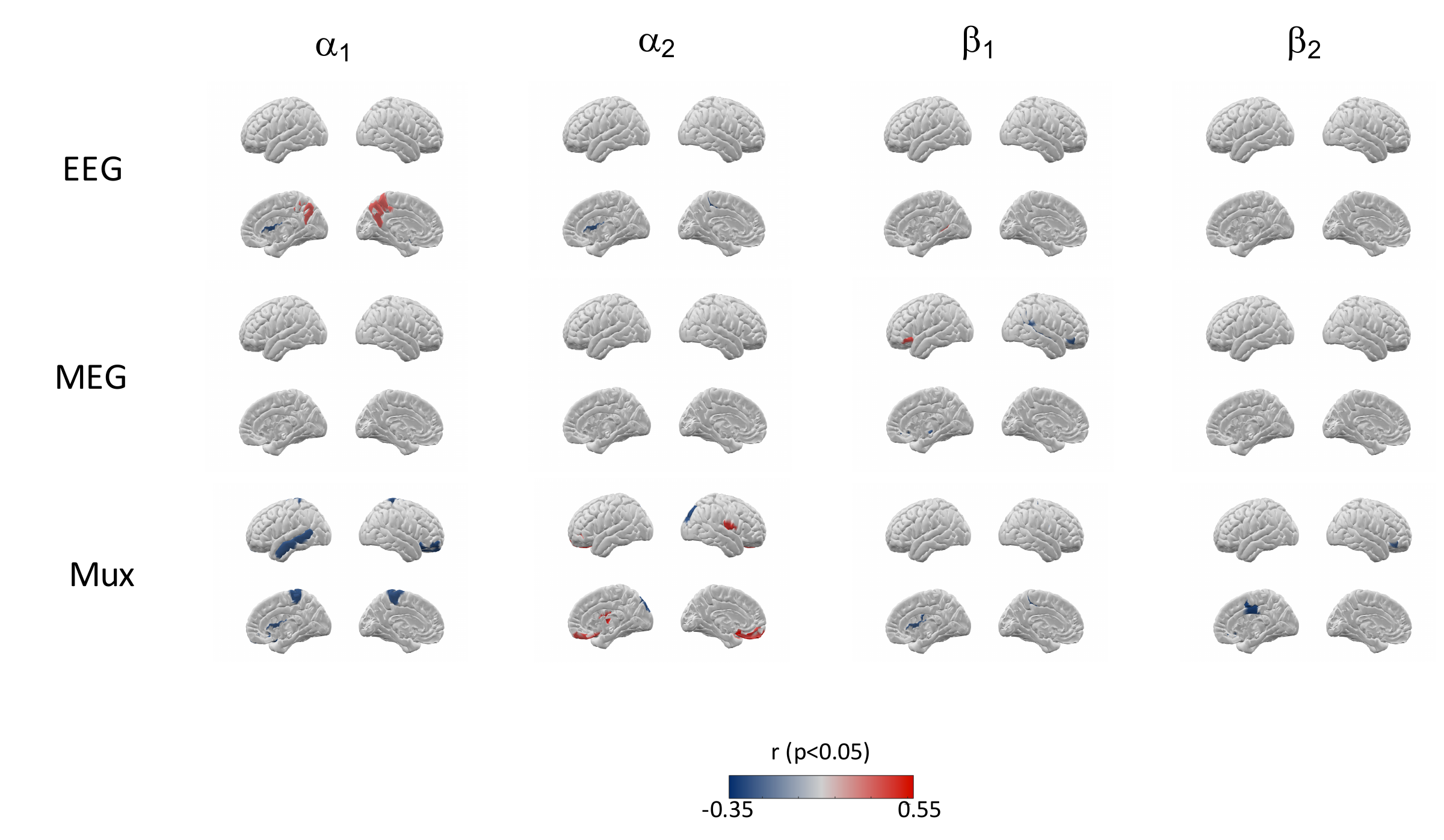} 
\caption{Repeated correlations between relative coreness and BCI scores obtained in the subsequent session in EEG, MEG, and mux ($p < 0.05$) within the $\alpha$ and the $\beta$ frequency ranges.}
\label{Figure11}
\end{figure}
\newpage

\clearpage

\section*{Supplementary Tables}

\begin{table}[!ht]
%\begin{fullwidth}
\caption{List of ROIs that show a significant condition effect at least once during the training in the EEG modality ($p<0.021$). In $\beta_{2}$, we did not observe any significant result.} 
\centering
\resizebox{\textwidth}{!}{
\begin{tabular}{cccc rrrrrr rr} 
\hline
Frequency bands & ROI & (t, p)  & Session \\
\hline
 \multirow{9}{*}{$\alpha_{1}$} & Anterior part of the cingulate gyrus and sulcus L & (3.02	, 0.007) & S3 \\ 
& Anterior part of the cingulate gyrus and sulcus R & (2.55, 0.020) & S4\\  
& Middle-Anterior part of the cingulate gyrus and sulcus L & (2.92, 0.009) & S4 \\  
& Opercular part of the inferior frontal gyrus R & (2.81, 0.011) & S4 \\ 
& Lateral occipito-temporal gyrus R & (2.73, 0.013) & S2 \\
& Superior occipital gyrus L & (-2.54, 0.020) & S1  \\ 
& Precuneus L & (-2.37, 0.028; -3.79, 0.001) & (S1; S2) \\
& Planum polare of the superior temporal gyrus L & (3.13,	0.005) & S4 \\
& Vertical ramus of the anterior segment of the lateral sulcus R & (2.66, 0.016) & S4 \\
\hline
\multirow{10}{*}{$\alpha_{2}$} & Long insular gyrus and central sulcus of the insula L & (3.28, 0.004; 2.57, 0.019) & (S3; S4) \\
& Long insular gyrus and central sulcus of the insula R & (3.69,	0.002) & S3 \\
& Middle-Anterior part of the cingulate gyrus and sulcus L & (2.72, 0.014) & S4 \\  
& Middle-posterior part of the cingulate gyrus and sulcusL & (2.81, 0.011) & S3 \\ 
& Orbital part of the inferior frontal gyrus R & (2.58, 0.018) & S4 \\  
& Lingual gyrus R & (2.58,	0.018) & S4 \\  
& Superior parietal lobule L & (-3.46	0.003) & S1 \\
& Postcentral gyrus R & (-2.29, 0.034) & S2 \\
& Gyrus rectus L & (2.61, 0.017) & S3 \\
& Gyrus rectus R & (3.05, 0.007) & S4 \\
& Posterior ramus (or segment) of the lateral sulcus (or fissure) R & (3.07,	 0.006) & S1 \\
& Temporal pole R & (3.18, 0.005) & S1 \\
 \hline \hline 
 \multirow{7}{*}{$\beta_{1}$} 
 & Transverse frontopolar gyri and sulci R & (2.71, 0.014) & S2  \\
 & Subcallosal gyrus R & (2.65, 0.016) & S3  \\
 & Anterior transverse temporal gyrus L & (2.60, 0.017) & S3  \\
 & Planum polare of the superior temporal gyrus R & (3.16, 0.005) & S2  \\
 & Inferior temporal gyrus L & (3.53, 	0.002) & S4  \\
 & Middle temporal gyrus L & (3.02, 0.007) & S4  \\
 & Posterior ramus (or segment) of the lateral sulcus (or fissure) R & (3.29, 0.004) & S2  \\
 \hline
\label{Table1}
\end{tabular}
}
%\end{fullwidth}
\end{table}
\newpage

\begin{table}[bt]
%\begin{fullwidth}
\caption{List of ROIs that show a significant condition effect at least once during the training in the MEG modality ($p<0.021$).} 
\centering
\resizebox{\textwidth}{!}{
\begin{tabular}{ccccc rrrrrr rr} 
\hline
Frequency bands & ROI & (t, p)  & Session \\
\hline
 \multirow{8}{*}{$\alpha_{1}$} 
& Paracentral lobule and sulcus L & (2.90,	0.009) & S2 \\  
& Cuneus L & (-3.19, 0.005) & S4 \\
& Opercular part of the inferior frontal gyrus L & (3.84, 0.001) & S2 \\
& Short insular gyri L & (2.59, 0.018) & S4 \\
& Supramarginal gyrus L & (3.07, 0.006) & S1  \\ 
& Anterior transverse temporal gyrus L & (2.71,	0.014) & S4 \\
& Horizontal ramus of the anterior segment of the lateral sulcus R & (2.68, 0.015) & S4 \\
& Vertical ramus of the anterior segment of the lateral sulcus L & (2.68, 0.015) & S2 \\
\hline
\multirow{4}{*}{$\alpha_{2}$} & Middle-Anterior part of the cingulate gyrus and sulcus L & (-2.70, 0.014) & (S2) \\
& Cuneus L & (-2.62, 0.017) & S3 \\ 
& Triangular part of the inferior frontal gyrus R & (2.89, 0.009) & S3 \\ 
& Supramarginal gyrus L & (2.59, 0.018) & S3 \\ 
 \hline \hline 
 \multirow{3}{*}{$\beta_{1}$} & Orbital part of the inferior frontal gyrus L & (2.64, 0.016) & S3 \\
& Orbital part of the inferior frontal gyrus R & (-2.57	0.019) & S3 \\
& Lateral occipito-temporal gyrus L & (3.27, 0.004) & S3 \\
 \hline
  \multirow{6}{*}{$\beta_{2}$} 
& Transverse frontopolar gyri and sulci R & (-2.86, 0.010) & S3 \\
& Posterior-ventral part of the cingulate gyrus L & (-3.10, 0.006) & S1 \\
& Orbital part of the inferior frontal gyrus R & (-2.67,	0.015) & S4 \\
& Lateral occipito-temporal gyrus L & (2.56,	0.019) & S3 \\
& Postcentral gyrus R & (-2.72,	0.014) & S2 \\
& Planum polare of the superior temporal gyrus R & (-2.64, 0.016) & S1 \\
   \hline
\label{Table2}
\end{tabular}
}
%\end{fullwidth}
\end{table}
\newpage

\begin{table}[bt]
%\begin{fullwidth}
\caption{List of ROIs that show a significant condition effect at least once during the training in the mux modality ($p<0.021$).} 
\centering
\resizebox{\textwidth}{!}{
\begin{tabular}{ccccc rrrrrr rr} 
\hline
Frequency bands & ROI & (t, p)  & Session \\
\hline
 \multirow{10}{*}{$\alpha_{1}$} & Anterior part of the cingulate gyrus and sulcus L & (2.64, 0.016) & S3 \\ 
& Middle-posterior part of the cingulate gyrus and sulcus L & (-2.71, 0.014) & S2 \\ 
& Transverse frontopolar gyri and sulci R & (2.86, 0.010) & S1 \\ 
& Posterior-dorsal part of the cingulate gyrus R & (-3.01	0.007) & S2 \\ 
& Opercular part of the inferior frontal gyrus L & (2.53	0.020) & S4 \\
& Superior occipital gyrus L & (-2.60, 0.017) & S1 \\ 
& Precuneus L & (-3.43,	0.003) & S2 \\ 
& Gyrus rectus L & (2.63, 0.017) & S4 \\ 
& Anterior transverse temporal gyrus L & (2.53, 0.020) & S4 \\ 
& Planum polare of the superior temporal gyrus L & (2.78	0.012) & S4 \\  
\hline
 \multirow{10}{*}{$\alpha_{2}$} 
& Long insular gyrus and central sulcus of the insula R & (3.09, 0.006) & S3 \\
& Middle-posterior part of the cingulate gyrus and sulcus L & (2.79, 0.012) & S4 \\
& Subcentral gyrus L & (2.54, 0.020) & S4 \\
& Cuneus R & (-2.60, 0.018) & S1 \\
& Superior parietal lobule L & (-2.87, 0.010) & S2 \\
& Precuneus L & (-2.64, 0.016) & S2 \\
& Gyrus rectus L & (2.62, 0.017; 2.97, 0.008) & (S3; S4) \\
& Gyrus rectus R & (2.65, 0.016) & S3 \\
& Planum temporale or temporal plane of the superior temporal gyrus L & (2.37, 0.029) & S4 \\
& Middle temporal gyrus L & (-2.45, 0.024) & S2 \\
 \hline \hline 
 \multirow{5}{*}{$\beta_{1}$} 
& Orbital part of the inferior frontal gyrus R & (-3.41, 0.003) & S3 \\
& Short insular gyri R & (2.85, 0.010) & S1 \\
& Lateral occipito-temporal gyrus R & (2.56, 0.019) & S3 \\
& Gyrus rectus R & (2.72,	0.014) & S1 \\
& Planum polare of the superior temporal gyrus L & (2.72, 0.014) & S1 \\
 \hline
  \multirow{5}{*}{$\beta_{2}$} 
& Posterior-ventral part of the cingulate gyrus L & (2.76, 0.012) & S2 \\
& Posterior-ventral part of the cingulate gyrus R & (3.33,	0.004) & S2 \\
& Opercular part of the inferior frontal gyrus L & (-2.62, 0.017) & S1 \\
& Lateral occipito-temporal gyrus R & (2.89, 0.009) & S3 \\
& Occipital pole R & (2.58, 0.019) & S2 \\
   \hline
\label{Table3}
\end{tabular}
}
%\end{fullwidth}
\end{table}